\documentclass[aip,jcp,reprint,showpacs,longbibliography,noeprint]{revtex4-1} 
\usepackage[utf8]{inputenc}
\usepackage[english]{babel}
\usepackage{cmap} 
\usepackage[T1]{fontenc} 
\usepackage{multirow}
\usepackage{adjustbox}

\usepackage{amsmath,amssymb,bm,scalerel,mwe}
\usepackage{dsfont}
\IfFileExists{mtpro2.sty}
{ 
  \usepackage{times}
  \usepackage[lite,eucal,subscriptcorrection,slantedGreek,zswash]{mtpro2}
}{
\usepackage{txfonts}

}

\usepackage{mleftright}
\usepackage{microtype}
\usepackage{bbm}
\usepackage[title]{appendix}

\usepackage{graphicx,grffile,subfigure}
\graphicspath{{figures/}}
\newlength{\figwidth}
\usepackage{url,hyperref}
\usepackage[table,usenames,dvipsnames]{xcolor}
\hypersetup{colorlinks=true, linkcolor=BrickRed,
  urlcolor=blue!50!black, citecolor=blue!50!black}

\usepackage{xpatch}
\usepackage{appendix}
\usepackage[capitalize]{cleveref}
\crefrangelabelformat{equation}{(#3#1#4)$-$(#5#2#6)}
\let\ref\cref

\crefrangelabelformat{equation}{(#3#1#4)$-$(#5#2#6)}
\xapptocmd\appendices{%
  \crefalias{section}{appendix}%
}{}{\PatchFailed}

\usepackage{textcomp,yhmath,ushort}

\usepackage{tikz} 
\newcommand{\centered}[1]{\begin{tabular}{l} #1 \end{tabular}}

\newenvironment{frcseries}{\fontfamily{frc}\selectfont}{}

\newcommand{\sqdiamond}[1][fill=black]{\tikz
  [x=1.2ex,y=1.85ex,line width=.1ex,line join=round, yshift=-0.285ex]
  \draw [#1] 
  (-0.3,.5) -- (.5,1.) -- (1.3,.5) -- (.5,0) -- (-0.3,.5) -- cycle;}%

 \newcommand{\dotdashedline}[1][fill=black]{\tikz 
 [line width=.3ex]
 \draw [#1] [dash pattern=on 3pt off 0.5pt on \the\pgflinewidth off 0.5pt] 
 (0,0) -- (.835,0);}%

\newcommand{\MyDiamond}[1][fill=black]{\mathop{\raisebox{-0.275ex}{$\sqdiamond[#1]$}}}

\renewcommand{\bar}{\overline}
\newcommand{\bgma}{\boldsymbol{\Gamma}}
\newcommand{\bfet}{\boldsymbol{\eta}}
\newcommand{\barG}{\bar{\Gamma}}
\newcommand{\id}{\mathds{1}}
\newcommand{\bfR}{\mathbf{R}}
\newcommand{\matR}{\boldsymbol{\mathcal{R}}}

\renewcommand{\vec}[1]{\mathbf{#1}}

\newcommand{\kb}{k_{\rm B}}
\renewcommand{\vec}[1]{\mathbf{#1}}
\newcommand*\upd{\mathop{}\!\mathrm{d}}
\newcommand{\order}{\mathcal{O}}

\newcommand{\complexi}{\mathbbm{i}}

\newcommand\doverline[1]{\ThisStyle{%
  \setbox0=\hbox{$\SavedStyle\overline{#1}$}%
  \ht0=\dimexpr\ht0-.15ex\relax
  \overline{\copy0}%
}}

 \makeatletter
 \providecommand*{\diff}%
 	{\@ifnextchar^{\DIfF}{\DIfF^{}}}
    \def\DIfF^#1{%
    	\mathop{\mathrm{\mathstrut d}}%
    		\nolimits^{#1}\gobblespace
    }
    \def\gobblespace{%
    	\futurelet\diffarg\opspace}
    \def\opspace{%
    	\let\DiffSpace\!%
    	\ifx\diffarg(%
    		\let\DiffSpace\relax
    	\else
    		\ifx\diffarg\[%
    			\let\DiffSpace\relax
    		\else
    			\ifx\diffarg\{%
    				\let\DiffSpace\relax
    			\fi\fi\fi\DiffSpace}

\definecolor{myblue}{RGB}{94.3418,129.735, 181.708}
\definecolor{myokker}{RGB}{225.465, 156.426, 36.3651}
\definecolor{mygreen}{RGB}{143.406, 177.042, 49.8906}
\definecolor{myorange}{RGB}{236.167, 98.7203, 53.5498}
\definecolor{mypurple}{RGB}{135.293, 120.48, 179.546}
\definecolor{mybrown}{RGB}{197.652, 110.478, 26.2111}
\definecolor{mycyan}{RGB}{93.1579, 158.336, 200.281}
\definecolor{myyellow}{RGB}{255, 192, 0}
\definecolor{mymagenta}{RGB}{165.792, 96.809, 157.193}

\usepackage[normalem]{ulem}

\newcommand{\etal}{{\em et al.}}

\begin{document}
\def\bea{\begin{eqnarray}}
\def\eea{\end{eqnarray}}
\def\beq{\begin{equation}}
\def\eeq{\end{equation}}
\def\f{\frac}
\def\k{\kappa}
\def\e{\epsilon}
\def\ve{\varepsilon}
\def\be{\beta}
\def\D{\Delta}
\def\h{\theta}
\def\t{\tau}
\def\a{\alpha}

\def\rv{{\bf r}}
\def\jv{{\bf j}}
\def\kv{{\bf k}}
\def\Gv{{\bf G}}

\def\eff{{\rm eff}}
\def\mf{{\rm mf}}
\def\cDa{{\cal D}[X]}
\def\cD{{\cal D}[x]}
\def\cL{{\cal L}}
\def\cLo{{\cal L}_0}
\def\cLa{{\cal L}_1}

\def\Re{{\rm Re}}
\def\sj{\sum_{j=1}^2}
\def\rk{\rho^{ (k) }}
\def\rek{\rho^{ (1) }}
\def\cek{C^{ (1) }}
\def\rz{\rho^{ (0) }}
\def\rt{\rho^{ (2) }}
\def\rtb{\bar \rho^{ (2) }}
\def\trk{\tilde\rho^{ (k) }}
\def\trek{\tilde\rho^{ (1) }}
\def\trz{\tilde\rho^{ (0) }}
\def\trt{\tilde\rho^{ (2) }}
\def\r{\rho}
\def\tD{\tilde {D}}

\def\s{\sigma}
\def\kb{k_B}
\def\la{\langle}
\def\ra{\rangle}
\def\nn{\nonumber}
\def\up{\uparrow}
\def\dn{\downarrow}
\def\S{\Sigma}
\def\dg{\dagger}
\def\d{\delta}
\def\p{\partial}
\def\l{\lambda}
\def\L{\Lambda}
\def\G{\Gamma}
\def\o{\Omega}
\def\w{\omega}
\def\g{\gamma}
\def\upd{\rm d}
\def\noi{\noindent}
\def\a{\alpha}
\def\d{\delta}
\def\p{\partial} 

\def\la{\langle}
\def\ra{\rangle}
\def\e{\epsilon}
\def\n{\eta}
\def\g{\gamma}
\def\break#1{\pagebreak \vspace*{#1}}
\def\hf{\frac{1}{2}}
\def\bgma{\boldsymbol{\Gamma}}
 
\title{Persistence in Brownian motion of an ellipsoidal
  particle in two dimensions} 
\author{Anirban Ghosh} 
\affiliation{Indian Institute of Science Education and Research Mohali,
  Sec. 81, S.A.S. Nagar, Knowledge City, Manauli, Punjab-140306,
  India.}
\author{Dipanjan Chakraborty} 
\affiliation{Indian Institute of Science Education and Research Mohali,
  Sec. 81, S.A.S. Nagar, Knowledge City, Manauli, Punjab-140306, India.}
\email{chakraborty@iisermohali.ac.in}

\date{\today}
\begin{abstract}
  We investigate the persistence probability $p(t)$ of the position of
  a Brownian particle with shape asymmetry in two dimensions. The
  persistence probability is defined as the probability that a
  stochastic variable has not changed it's sign in the given time
  interval. We explicitly consider two cases -- diffusion of a free
  particle and that of harmonically trapped particle. The later is
  particularly relevant in experiments which uses trapping and
  tracking techniques to measure the displacements. We provide
  analytical expressions of $p(t)$ for both the scenarios and show
  that in the absence of the shape asymmetry the results reduce to the
  case of an isotropic particle. The analytical expressions of $p(t)$
  are further validated against numerical simulation of the underlying
  overdamped dynamics. We also illustrate that $p(t)$ can be a measure
  to determine the shape asymmetry of a colloid and the translational
  and rotational diffusivities can be estimated from the measured
  persistence probability. The advantage of this method is that it
  does not require the tracking of the orientation of the particle.
\end{abstract}
\maketitle
\section{Introduction}
\label{sec:intro}
Particles that exhibit a shape asymmetry are abundant in nature with
sizes ranging from few nanometers to few micrometers. Over the last
decade, accelarated by the advancement in particle chemistry, a
plethora of such particles with enhanced transport properties have
been developed in an attempt to mimic nature. These synthetically
engineered colloids with multi-functional properties often find wide
ranging applications in photonics, nano and biotechnology, drug
delivery and other bio-medical uses. Unlike an isotropic particle, the
shape asymmetry leads to different transport properties along the
symmetry axes of the particle and any real-life application would
require the knowledge of these transport properties. Perhaps, the most
crucial of these transport properties are the translational and
rotational diffusivities that characterizes their stochastic dynamics.
For example, the diffusive dynamics of such particles are completely
characterized by the mobility matrix. However, the extraction of the
diffusivity from the measured mean-square displacement requires
the simultaneous measurement of its translational and orientational
degrees of freedom, which might not be always feasible. 

In this article, we present an alternative approach to measure the
diffusivity of shape asymmetric particle from its position coordinates
alone. Our approach does not require the measurement of the symmetry
axes of the particle. We choose the simplest asymmetric particle -- an
ellipsoid and look at its two dimensional Brownian motion. Since the
dynamics of the translational and the orientational degrees of freedom
are stochastic due to the thermal fluctuations from the bath, the
position and the orientation are both random variables in time.  We
use the stochastic nature of the position to calculate the persistence
probability $p(t)$ of the particle. The extraction of the diffusion
coefficients along the two symmetry axes of the particle as well the
rotational diffusion constant follows from the analytical expression
of $p(t)$.

The persistence probability $p(t)$ of a stochastic variable is simply
the probability that the variable has not changed sign up to time
$t$. In physics, the persistence property has been investigated both
theoretically
\cite{derrida1995,newman1998,kallabis1999,toroczkai1999,sire2000,constantin2004,bray2004a,majumdar1996,majumdar1996b,majumdar1999,majumdar2001,
  chakraborty2007,chakraborty2007c,chakraborty2008,chakraborty2009,chakraborty2012c,chakraborty2012d,constantin2005,dean2001a,
escudero2009,menon2003,ray2004a,krug1997,singha2005}
and experimentally \cite{Wong:2001dr,dougherty2002,Merikoski:2003ju,
  Beysens:2006jt,Soriano:2009br,Efraim:2011ks,Takeuchi:2012fe,takikawa2013}
in spatially extended systems that are out of
equilibrium.  
For a more comprehensive review of the persistence probability in
spatially extended systems, we invite the readers to look at the
recent review by Bray \etal \cite{bray2013b} and the brief review by
Majumdar \cite{Majumdar:1999tn} on the subject and the references
therein.  The persistence probability for such systems decays as a
power law $p(t) \sim t^{-\theta}$, with $\theta$ being a non-trivial
exponent.  This algebraic decay of $p(t)$ has been established for a
wide class of non-equilibrium systems that includes the classic random
walk problem in finite \cite{chakraborty2007} and infinite medium
\cite{Majumdar:1999tn,sire2000,bray2004a,chakraborty2012d}, critical
dynamics \cite{majumdar1996b,chakraborty2007c}, diffusion in an
infinite medium with \cite{chakraborty2009} and without
advection\cite{majumdar1996,newman1998}, fluctuating interfaces
\cite{derrida1995,krug1997,kallabis1999,toroczkai1999,constantin2004},
disordered systems
\cite{Fisher:1998km,LeDoussal:1999ik,chakraborty2008}, polymer
dynamics \cite{bhattacharya2007,chakraborty2012c} and granular media
\cite{Swift:1999kn,Burkhardt:2000hs}.
The estimation of the exponent $\theta$ for a general stochastic
process is notoriously difficult and the exact form of $p(t)$ exists
in very few cases when the process is Gaussian as well Markovian.  For
a stochastic process $x(t)$ which is Gaussian as well as Markovian,
the non-stationary process can be mapped into a stationary
Ornstein-Uhlenbeck process $\bar{X}(T)$ via suitable transformations
that takes $x \to \bar{X}$ and $t \to T$, with the consequence that
the correlator $C(T) \equiv \langle \bar{X}(T) \bar{X}(0)\rangle$
decay exponentially at all times.Following Slepian \cite{slepian1962},
if the stationary correlator $C(T)$ of a stochastic process decays
purely exponentially at all times, the persistence probability of
$X(T)$ is proportional to $C(T)$ and $p(t)$ can then be constructed
back by the inverse time transformation applied to $\bar{X}$. In the
case when $C(T)$ does not decay exponentially, the exponent $\theta$
can be extracted using the independent interval approximation (IIA) ,
provided the density of zero crossings remain
finite\cite{majumdar1996}. In the present scenario, as the
calculations reveal, the IIA is not required and suitable
transformations space and time takes the non-stationary correlation
function into a stationary correlator which then be used to calculate
$p(t)$.

The rest of the article is organized as follows. In
\cref{sec:free_ellipsoidal_particle} present the results for the
two-time correlation function the position of a free Brownian particle
with shape asymmetry. The survival probability is determined from this
correlation function. In \cref{sec:harmonically_trapped} we carry out
a perturbative expansion for the position of an anisotropic Brownian
particle trapped in a harmonic potential. The mean-square displacement
for the displacements along the two directions and the two-time
correlation functions are calculated using the perturbative
expansion. Finally, the persistence probability is constructed from
this two-time correlation function. A brief conclusion and the
relevance of the work is presented in \cref{sec:conclusion}.

\section{Ellipsoidal Particle in two-dimensions}
\label{sec:free_ellipsoidal_particle}

We consider an ellipsoidal particle in two dimension with mobilities 
$\Gamma_x$ and $\Gamma_y$ along the $x$ and $y$ direction respectively and 
a single rotational mobility $\Gamma_\theta$. The particle
is immersed in a bath at a temperature $T$, so that the translational 
diffusion coefficients along the two directions are given by $D_x=\kb T 
\Gamma_x$, $D_y =\kb T \Gamma_y$ and the rotational diffusion
constant $D_\theta =\kb T \Gamma_\theta$. In a frame fixed to the 
particle, the translational and the rotational motion of the particle is 
completely decoupled.However, in the lab-frame, the shape asymmetry of the 
particle leads to a coupling between the translational and rotational 
motions of the particle. In the body frame the equations of motion of the 
particle take the form
\begin{align}
  \label{eq:body_frame}
  \Gamma_x^{-1}\frac{\partial \tilde{x}}{\partial t}=&F_x \cos
  \theta(t) +F_y \sin \theta(t) +\tilde{\eta}_x(t) \nonumber\\
  \Gamma_y^{-1}\frac{\partial \tilde{y}}{\partial t}=&F_y \cos
  \theta(t) +F_x \sin \theta(t) +\tilde{\eta}_y(t)\nonumber\\
\Gamma_\theta^{-1}\frac{\partial \theta}{\partial t}=& \tau+\tilde{\eta}_\theta,	
\end{align}
where $F_x$ and $F_y$ are the forces acting on the particle along 
the $x$ and $y$ directions and $\tau$ is the torque acting on
the particle. The correlations of the thermal fluctuations in the 
body frame are given by
\begin{equation}
\label{eq:noise_body_frame}
	\langle \tilde{\eta} \rangle=0 \quad \textrm{and} \quad  \langle \tilde{\eta}_i(t) \tilde{\eta}_j(t') \rangle=2 D_i \delta_{ij} \delta(t-t')
\end{equation}


In the lab frame, the displacements are related to the
body frame as,
\begin{eqnarray}
  \label{eq:lab_body}
\nonumber
  \delta x= \cos \theta \delta \tilde{x}-\sin \theta \delta \tilde{y}\\
  \delta y =\cos \theta \delta \tilde{y}+\sin \theta \delta \tilde{x}
\end{eqnarray}

Using \cref{eq:body_frame}, the corresponding Langevin equation in
the lab frame is given by,
\begin{equation}
  \label{eq:langevin}
  \frac{\partial x_i}{\partial t}= -\Gamma_{ij} \frac{\partial
    U}{\partial x_j} +\eta_i,
\end{equation}
where $U(\vec{r})$ is the external potential and 
$\boldsymbol{\Gamma}$ is the mobility tensor given by,
\begin{equation}
  \label{eq:mobility_tensor}
  \doverline{\bgma}=
\begin{pmatrix}
\barG+\frac{\D\Gamma}{2} \cos 2\theta & \frac{\Delta
  \Gamma}{2} \sin 2 \theta \\
\\
 \frac{\D\Gamma}{2} \sin 2 \theta & \barG-
\frac{\D\Gamma}{2} \cos 2\theta\\
\end{pmatrix}
\end{equation}
with $\barG=(\Gamma_{\parallel}+\Gamma_{\bot})/2$ and $\D\Gamma
=\Gamma_{\parallel}-\Gamma_{\bot}$. In the component form, the mobility tensor is
given by $\Gamma_{ij}=\barG\delta_{ij}+\frac{\D\Gamma}{2} \Delta
\matR_{ij}[\theta(t)]$, where the form of $\D
\doverline{\matR}$ is given by 
\begin{equation}
  \label{eq:delta_gamma_matrix}
\D\doverline{\matR}=
\begin{pmatrix}
\cos 2\theta & \sin 2 \theta \\
\\
 \sin 2 \theta & -\cos 2\theta\\
\end{pmatrix}
\end{equation}

Using the correlation of the thermal fluctuations from 
\cref{eq:noise_body_frame} and \cref{eq:body_frame}, 
the moments of the stochastic forces are given by,
\begin{equation}
  \label{eq:fdt}
  \la \bfet \ra =0 \quad \textrm{and} \quad \la
  \bfet(t)  \bfet(t') \ra= 2 \kb T
  \boldsymbol{ \Gamma}[\theta(t)] \delta(t-t') 
\end{equation}

We first look at the case of a free ellipsoidal particle. Setting 
the external potential to zero, the formal solution to the equation of 
motion takes the form 
\begin{equation}
\label{eq:free_particle_solution}
	x_i(t)=\int_0^t \eta_i(t') \mathrm{d}  t' + x_i(0)
\end{equation}
The mean-square displacement of the particle, averaged over the 
orientational noise can be explicitly calculated from the above 
equation as,
	\begin{align}
		\label{eq:msd}
	\la \Delta x_i^2\ra_{\eta_\theta} &=\int_0^t \mathrm{d} t' \int_0^t {\diff} t'' \la \eta_i(t')\eta_i(t'')\ra \nonumber\\
	&=2 \kb T \int_0^t \mathrm{d} t' \int_0^t {\diff} t'' ~~\la \Gamma_{ii}[\theta(t')]\ra_{\eta_\theta}
	\delta(t'-t'') \nonumber\\
	&=2 \kb T \int_0^t \mathrm{d} t'  ~~\la \Gamma_{ii}[\theta(t')]\ra_{\eta_\theta}		
	\end{align}
Using the explicit form of $\Gamma_{xx}$ the mean-square
displacement along the $x$-direction reads
\begin{equation}
\label{eq:xsq_free}
\la \Delta x_1^2\ra_{\eta_\theta} =2\kb T \int_0^t {\diff} t' ~~\left[ \bar{\Gamma}+ \frac{\Delta \Gamma}{2} \la \cos \theta(t') \ra_{\eta_\theta}\right]	
\end{equation}
The ensemble average of $\cos \theta(t)$ over the thermal fluctuations
in the orientational degrees of freedom can be done explicitly by
noting the fact that $\Delta \theta=\theta(t)- \theta_0$ is a Gaussian
random variable and consequently the following identity holds:
\begin{equation}
\label{eq:identity_1}
\la e^{\pm\complexi m \Delta \theta(t')}\ra_{\eta_\theta}=
e^{-m^2 D_\theta t'}.
\end{equation}
Using \cref{eq:identity_1} in \cref{eq:xsq_free}, we finally arrive at
\begin{equation}
\label{eq:xsq_free_1}
\la \Delta x^2\ra_{\eta_\theta} =2\kb T ~~\left[ \bar{\Gamma} t+ \frac{\Delta \Gamma}{2} \cos 2\theta_0 \left(\frac{1-e^{-4 D_\theta t}}{4 D_\theta}\right)\right]	
\end{equation}
and 
\begin{equation}
\label{eq:ysq_free_1}
\la \Delta y^2\ra_{\eta_\theta} =2\kb T ~~\left[ \bar{\Gamma} t- \frac{\Delta \Gamma}{2} \sin 2\theta_0 \left(\frac{1-e^{-4 D_\theta t}}{4 D_\theta }\right)\right]	
\end{equation}
The above results are well known \cite{Han:2006ew,Grima:2007fo} and
have also been experimentally verified. \cite{Han:2006ew} However, our
interest lies in the persistence probability of this system. To
calculate that we start with \cref{eq:free_particle_solution} and
choose $x_i(0)=0$. The calculation of the two time correlation
function $\la x(t_1) x(t_2) \ra_{\eta_\theta}$ follows the same route
as detailed above:
\begin{equation}
	\la x(t_1) x(t_2) \ra_{\eta_\theta} = \int_0^{t_1} \mathrm{d}t' \int_0^{t_2} \mathrm{d} t'' \la \eta_x(t') \eta_x(t'')\ra_{\eta_\theta}.
\end{equation}
Taking $t_1>t_2$, the integral evaluates to the following expression for the two time correlation,
\begin{equation}
	\la x(t_1) x(t_2) \ra_{\eta_\theta} = 2 \kb T \bar{\Gamma} t_2 \left[ 1+ \frac{\Delta \Gamma}{2\bar{\Gamma}}  \cos \theta_0 \left(\frac{1-e^{-4D_\theta t_2}}{4 D_\theta t_2} \right) \right]
\end{equation}
In order to transform the non-stationary correlation into a stationary
correlation we first make the transformation
$\bar{X}(t)=x(t)/\sqrt{\la x^2(t) \ra_{\eta_\theta}}$, and the
correlation $\la \bar{X}(t_1)\bar{X}(t_2)\ra_{\eta_\theta}$ reads as
\begin{equation}
\label{eq:corr_space}
\la \bar{X}(t_1)\bar{X}(t_2)\ra_{\eta_\theta}=\sqrt{\frac{2 \bar{D} t_2}{2 \bar{D} t_1}}\sqrt{\frac{1 +\frac{\Delta \Gamma}{2\bar{\Gamma}} \cos \theta_0\left(\frac{1-e^{-4D_\theta t_2}}{4 D_\theta t_2} \right)}{1 + \frac{\Delta \Gamma}{2 \bar{\Gamma}} \cos \theta_0 \left(\frac{1-e^{-4D_\theta t_1}}{4 D_\theta t_1} \right)}}
\end{equation}
We now define the transformation in time 
as 
\begin{equation}
\label{eq:time_transformation}	
e^T=\sqrt{2 \bar{D} t \left[1 +\frac{\Delta D}{2\bar{D}} \cos \theta_0\left(\frac{1-e^{-4D_\theta t}}{4 D_\theta t} \right)\right]}
\end{equation}
and \cref{eq:corr_space} takes the simple form of 
\begin{equation}
\label{eq:corr_space_time}
	\la \bar{X}(T_1)\bar{X}(T_2)\ra_{\eta_\theta}=e^{-(T_1-T_2)/2}
\end{equation}
Following Slepian \cite{slepian1962}, if the correlation function of a stochastic variable $X(T)$ 
decays exponentially for all times $C_{XX}(T) = e^{-\lambda T}$, then the persistence probability is given by
\begin{equation}
	\label{eq:per_prob}
	P(T) \sim \sin^{-1} e^{-\lambda T}.
\end{equation}
Asymptotically, $P(T)$ takes the form $P(T) \sim e^{-\lambda
  T}$.
Consequently, looking at \cref{eq:corr_space_time} and transforming
back in real time $t$, the persistence probability reads as
\begin{equation}
	\label{eq:per_prob_real_time}
	p(t) \sim \frac{1}{\sqrt{2\bar{D}
            t}}\frac{1}{\sqrt{1+\frac{\Delta D}{2 \bar{D}} \cos
            \theta_0  \left(\frac{1-e^{-4D_\theta t}}{4 D_\theta t} \right)}}.
\end{equation}
In the absence of any asymmetry, the expression for $p(t)$ correctly
reproduces the persistence probability of that of a random
walker. Rearranging \cref{eq:per_prob_real_time}, the quantity
$t^{1/2}p(t)$ can be recast as
\begin{equation}
  \label{eq:per_prob_real_time_1}
  t^{1/2} p(t) \sim \frac{1}{\sqrt{2 \bar{D}}}\left[1+\frac{\Delta D}{ \bar{D}} \cos \theta_0 \left(\frac{1-e^{-4 \tau}}{8 \tau} \right)\right]^{-1/2}.
\end{equation}
In the limit of $\D D \to 0$, the persistence probability reduces to
that of a random walker $p(t) \sim t^{-1/2}$.

\begin{figure}[!t]
  \centering
  \includegraphics[width=0.8\linewidth]{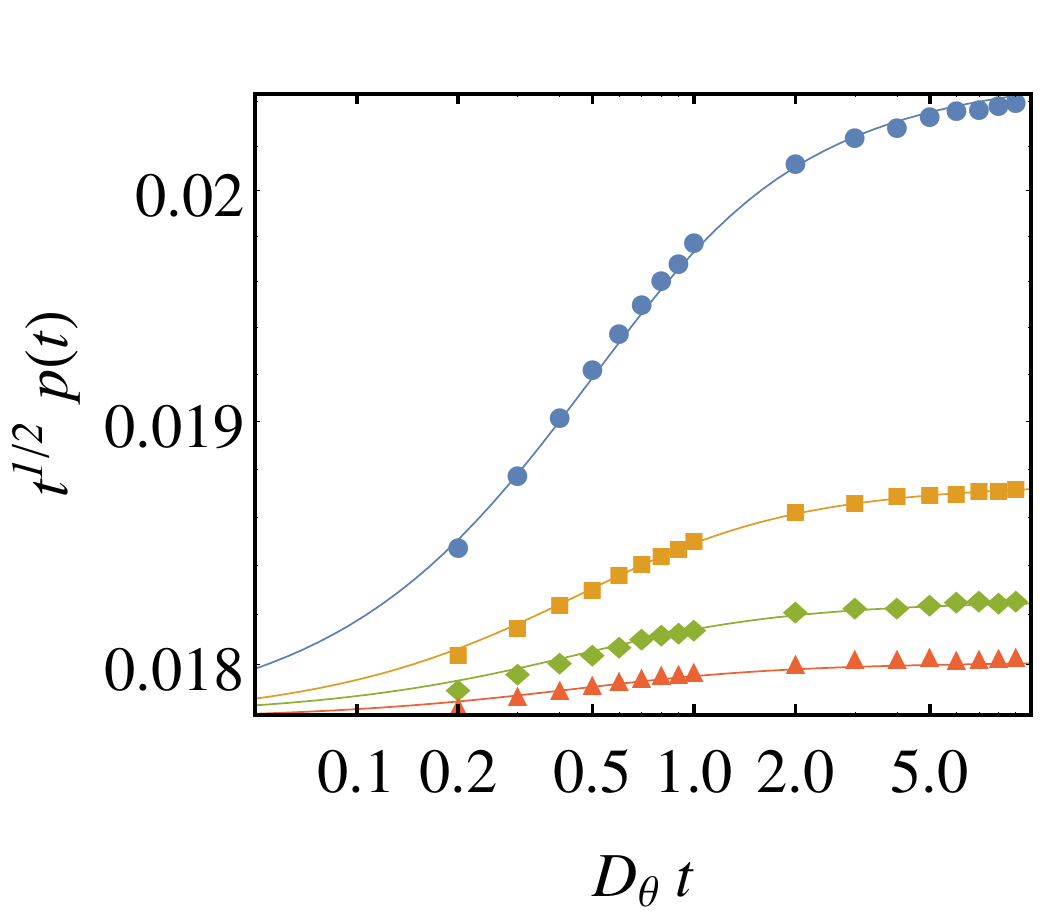}
  \caption{Plot of $t^{1/2}p(t)$ for different choices of
    translational diffusivities of the anisotropic particle: $D_{\parallel}=1,
    D_{\bot}=0.5$ ({\color{myblue} \Large $\bullet$}), $D_{\parallel}=1,
    D_{\bot}=0.8$ ({\color{myokker}
      $\blacksquare$}),$D_{\parallel}=1,
    D_{\bot}=0.9$  ({\scriptsize $\MyDiamond[draw=mygreen,fill=mygreen]$})  and
    $D_{\parallel}=1,
    D_{\bot}=0.95$ ({\color{myorange} $\blacktriangle$}). The
    rotational diffusion constant and the initial angle $\theta_0$ was
    fixed at $D_\theta=1$ and $\theta_0=0$,respectively. The solid lines
    are fit to the data using \cref{eq:per_prob_real_time_1}. The fit
    yields the overall constant $\mathcal{A}$. The estimated of values
    of $\mathcal{A}$ from the fit are $0.025132 \pm 0.000014$ for  $D_{\parallel}=1,
    D_{\bot}=0.5$, $0.025144 \pm 0.000011$ for  $D_{\parallel}=1,
    D_{\bot}=0.8$, $0.025166 \pm 0.000012$ for  $D_{\parallel}=1,
    D_{\bot}=0.9$ and $0.025148 \pm 0.000019$ for  $D_{\parallel}=1,
    D_{\bot}=0.95$. 
 }
  \label{fig:persistence_fig_1}
\end{figure}

\begin{figure}[!t]
  \centering
  \includegraphics[width=0.8\linewidth]{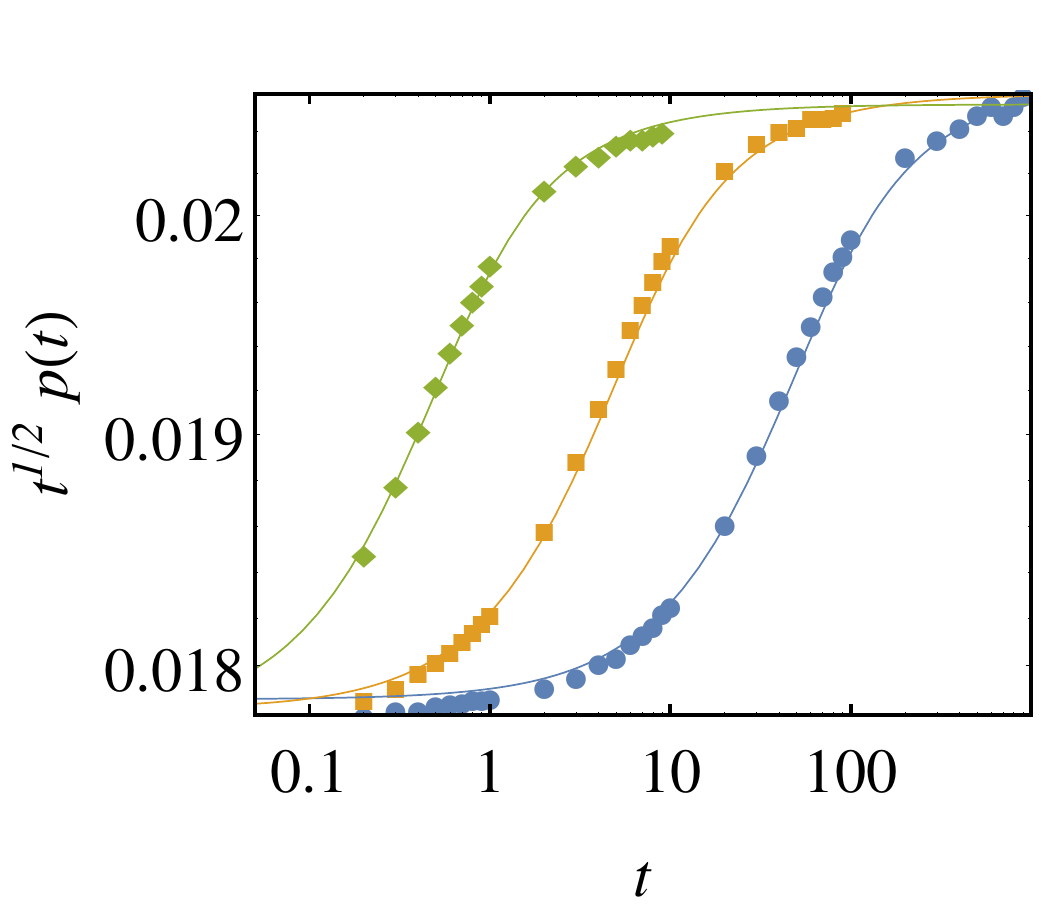}
  \caption{Plot of $t^{1/2}p(t)$ for different choices of
    rotational diffusion constant of the anisotropic particle:
    $D_\theta=0.01$ ({\color{myblue} \Large $\bullet$}),$D_\theta=0.1$
    ({\color{myokker} $\blacksquare$}),$D_\theta=1$ ({\scriptsize
      $\MyDiamond[draw=mygreen,fill=mygreen]$}). The translation
    diffusion constants in all the three cases were $D_{\parallel}=1$
    and $D_{\bot}=0.5$ and the initial orientation was fixed at $\theta_0=0$.}
  \label{fig:persistence_fig_2}
\end{figure}

To test \cref{eq:per_prob_real_time}, we performed numerical
integration of the equations of motion using an Euler scheme for
discritization. The initial condition was chosen from a Gaussian
distribution with a very small width, so that the sign of $\vec{r}(0)$
is clearly defined. The trajectories was evolved in time with an
integration time-step of $\delta t=0.001$. At every instant the the
survival of the particle was checked by looking at the sign of
$\vec{r}(t)$. Fraction of trajectories for which the position did not
change its sign up to time $t$ gave the survival probability $p(t)$. A
total of $10^9$ trajectories were used in estimating the survival
probability. A comparison of the measured $p(t)$ with that of the
predictions of \cref{eq:per_prob_real_time} is shown
\cref{fig:persistence_fig_1} and \cref{fig:persistence_fig_2}.
The comparison in \cref{fig:persistence_fig_1} clearly shows that
the survival probability can pick up the asymmetry in particle shape even
when the difference in the diffusivities is as small as $5\%$.


\begin{figure}[!t]
  \centering
  \includegraphics[width=0.8\linewidth]{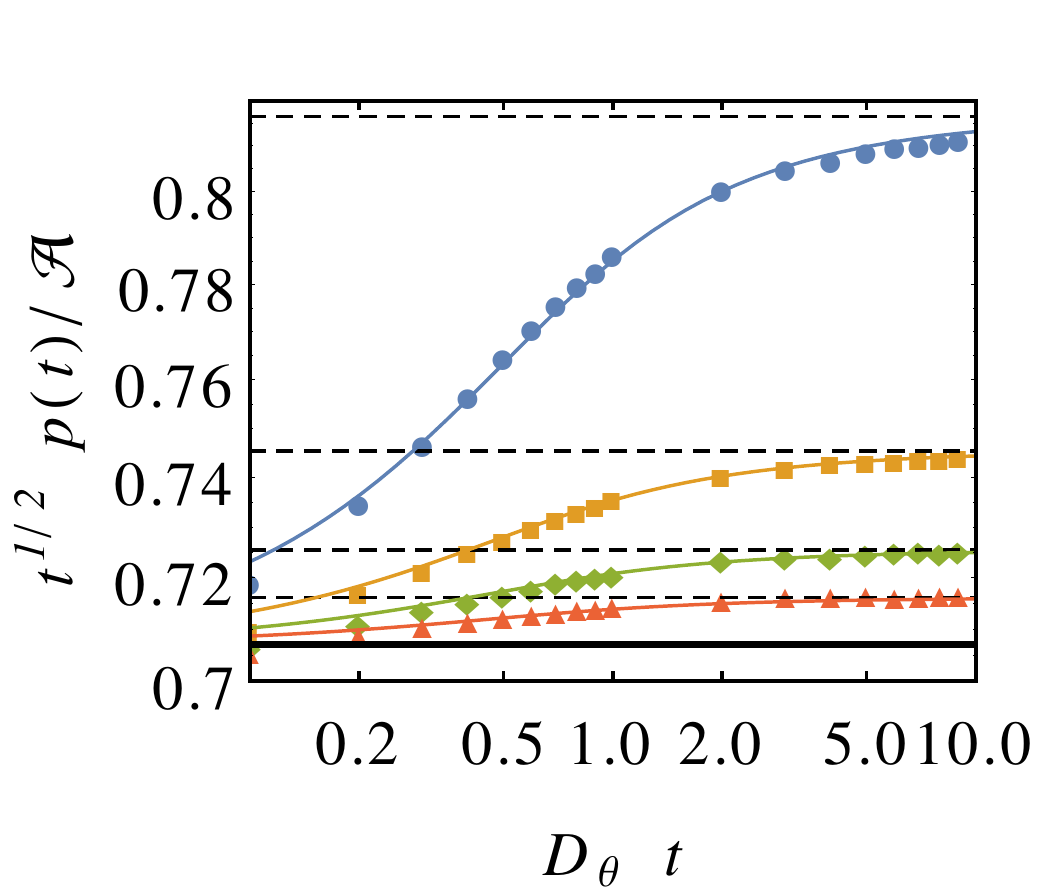}
  \caption{Plot of $t^{1/2}p(t)/\mathcal{A}$ for different choices of
    translational diffusivities of the anisotropic particle: $D_{\parallel}=1,
    D_{\bot}=0.5$ ({\color{myblue} \Large $\bullet$}), $D_{\parallel}=1,
    D_{\bot}=0.8$ ({\color{myokker}
      $\blacksquare$}),$D_{\parallel}=1,
    D_{\bot}=0.9$  ({\scriptsize $\MyDiamond[draw=mygreen,fill=mygreen]$})  and
    $D_{\parallel}=1,
    D_{\bot}=0.95$ ({\color{myorange} $\blacktriangle$}). The
    rotational diffusion constant and the initial orientation was
    fixed at $D_\theta=1$ and $\theta_0=0$, respectively. The solid black
    line indicates the value of $1/\sqrt{2\bar{D}+\Delta D}=1/\sqrt{2
      D_\parallel}=1/\sqrt{2}$, whereas the dashed lines indicates the
    values of $1/\sqrt{2 \bar{D}}$. For the choice of the translational
    diffusivities, the indicated values from top are $1/\sqrt{2
      \bar{D}} \approx 0.8165, 0.7454, 0.7255$ and $0.7161$.}
  \label{fig:persistence_fig_3}
\end{figure}

\begin{figure}[!t]
  \centering
  \includegraphics[width=0.8\linewidth]{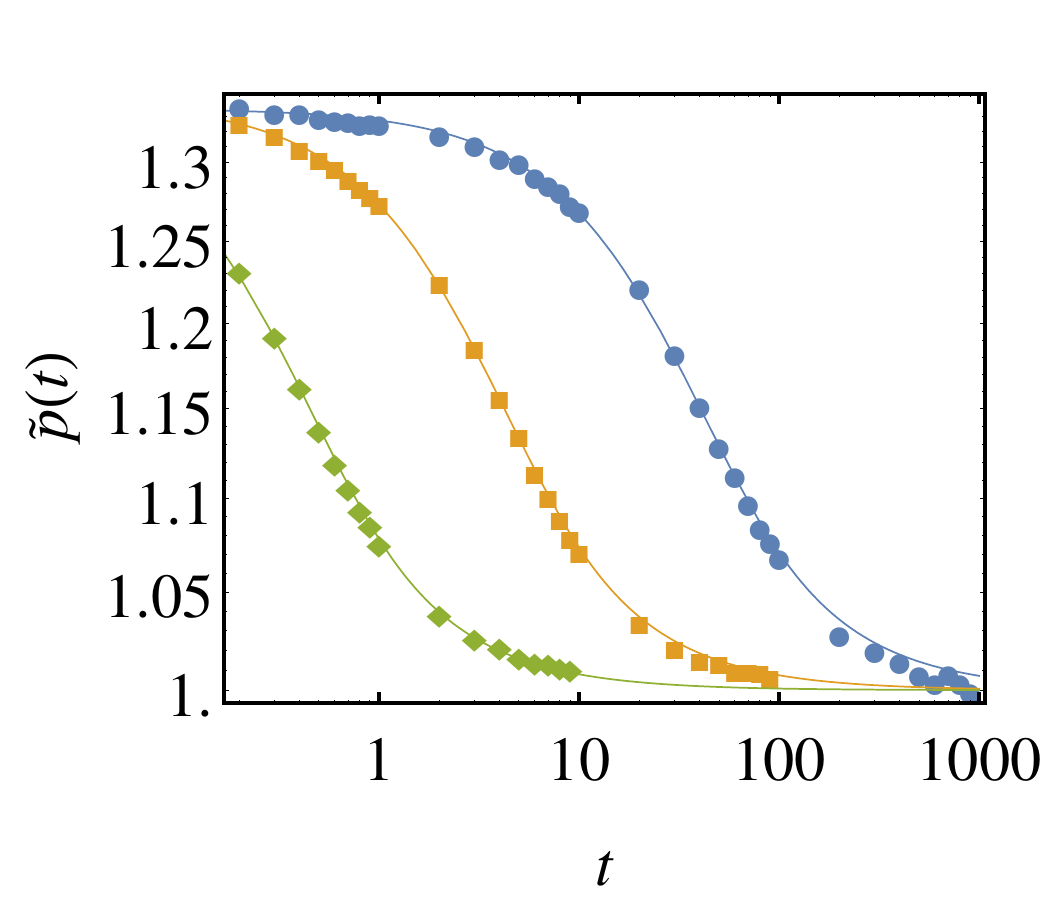}
  \caption{Plot of the dimensionless quantity $\tilde{p}(t)$ (see
    \cref{eq:rot_diff}) as function of time for different choices of
    rotational diffusion constant of the anisotropic particle:
    $D_\theta=0.01$ ({\color{myblue} \Large $\bullet$}),$D_\theta=0.1$
    ({\color{myokker} $\blacksquare$}),$D_\theta=1$ ({\scriptsize
      $\MyDiamond[draw=mygreen,fill=mygreen]$}). The translation
    diffusion constants in all the three cases were $D_{\parallel}=1$
    and $D_{\bot}=0.5$. The initial orientation in all the cases were
    fixed at $\theta_0=0$. The solid lines are fit to the data using
    \cref{eq:rot_diff} using $\Delta D/\bar{D}$ and $D_\theta$ as fit
    parameters. The estimated values of these parameters from the fit
    are compared with the actual values used in the simulation in
    \cref{tab:comparison_table}. }
  \label{fig:persistence_fig_4}
\end{figure}

The process to extract the the diffusion coefficients is as
follows. The first step would be to determine the overall constant
$\mathcal{A}$ in the expression for the persistence probability. This
can be fixed by fitting the data with the form of $p(t)$ given in
\cref{eq:per_prob_real_time}. This fit yields the value of
$\mathcal{A}$. In \cref{fig:persistence_fig_1}~(a), we have shown this
fitting for different choices of the diffusivities, with $\mathcal{A}$
as the fit parameter. The value of $\mathcal{A}$ is solely determined
by the number of trajectories used to estimate $p(t)$. The values
determined from the fit are given in the caption of the figure. An
alternative way to determine $\mathcal{A}$ is to measure the
persistence probability of an isotropic particle, in which case
$p(t) \sim \mathcal{A}/\sqrt{2 D t}$. Once this number is known, we
look at the quantity $t^{1/2} p(t)/\mathcal{A}$. In the limit of
$ t \to 0$, $t^{1/2} p(t)/\mathcal{A} \to (2 \bar{D}+\Delta D)^{1/2}$
and in the limit of $t \to \infty$,
$t^{1/2} p(t)/\mathcal{A} \to (2 \bar{D})^{1/2}$.

Once we know the two diffusivities, and therefore $\bar{D}$, the
rotational diffusion constant can be determined from the quantity
$\left(\mathcal{A}/\sqrt{2 \bar{D} t} p(t)\right)^2$ which goes as 
\begin{equation}
\label{eq:rot_diff}
\tilde{p}(t)=\left(\frac{\mathcal{A}}{\sqrt{2 \bar{D} t} p(t)}\right)^2=1+\left(\frac{\Delta D}{\bar{D}}\right)\left(\frac{1-e^{-4 D_\theta
  t}}{8D_\theta t}\right)
\end{equation}
A fit to $\tilde{p}(t)$ with $D_\theta$ as a fit parameter would yield
the value of the rotational diffusion coefficient. This is illustrated
in ~\cref{fig:persistence_fig_2}. In fact, fitting the data for $\tilde{p}$ with
$\Delta D/\bar{D}$ and $D_\theta$ as fit parameters yields very good
estimates for $\Delta D/\bar{D}$ and $D_\theta$. A comparison of these
values obtained from the fit with that of the actual values is shown
in \cref{tab:comparison_table}.


\begin{table}[h]
\centering
        \begin{tabular}{|*{4}{c|}}
          \hline
          \multirow{1}{*}{\centered{$\Delta D/\bar{D}$}}  &
                                                 \multirow{1}{*}{\centered{$D_\theta$}}
          &  \multirow{1}{*}{\centered{Estimated $\Delta D/\bar{D}$}}  &
                                                         \multirow{1}{*}{\centered{Estimated
                                                              $D_\theta$}}
          \\[8pt]
          \hline
          \multirow{3}{*}{$2/3$}  & 0.01 & $0.6698 \pm 0.0018$ &
                                                                 $0.0117
                                                                 \pm 0.0002$\\\cline{2-4}
                                  & 0.1 & $0.1146 \pm 0.0009$ &
                                                                $0.6799
                                                                \pm 0.002 $\\\cline{2-4}
                                  & 1.0 & $1.076 \pm 0.06$ & $0.681
                                                             \pm
                                                             0.0295$ \\\hline
        \end{tabular}
    \caption{A comparison of the actual values of $\Delta D/\bar{D}$
      and $D_\theta$ used in the simulations to those obtained from
      the fit of the data for $\mathcal{A}/2 \bar{D} t p^2(t)$.}
    \label{tab:comparison_table}
\end{table}

It should be pointed out, that the values of $\Delta D/\bar{D}$ and
$D_\theta$ obtained from the fit are sensitive to the value
$\mathcal{A}$ and a careful estimation of $\mathcal{A}$ is of
paramount importance.



\section{Harmonically trapped ellipsoidal particle}
\label{sec:harmonically_trapped}
In experiments, the tracking of colloidal particles are usually done
with laser traps and consequently it is pertinent to discuss the
scenario where an ellipsoidal particle is trapped in a harmonic
trap. In the following, we assume that the harmonic trap is isotropic
and there is no preferential direction of alignment. Further, if we
suppose a strong confinement, then at late times the deviations from
the mean position of the particle is practically zero. Accordingly,
the particle rotates freely so that the angular displacements obey
Gaussian statistics. The potential confinement has the form
$U(x,y)=\kappa (x^2+y^2)/2$ and the corresponding Langevin equation
from \cref{eq:langevin} take the form


\begin{align}
  \label{eq:langevin_harmonic}
  \frac{\partial x}{\partial t}=&-\kappa x \left(
                                  \bar{\Gamma}+\frac{1}{2}\Delta \Gamma \cos
  \theta(t)\right)-\frac{1}{2}\kappa y \Delta \Gamma \sin \theta(t) +\tilde{\eta}_x(t) \nonumber\\
  \frac{\partial y}{\partial t}=&-\frac{1}{2}\kappa x \Delta \Gamma \sin \theta(t)-\kappa y \left(
                                  \bar{\Gamma}-\frac{1}{2}\Delta \Gamma \cos
  \theta(t)\right) +\tilde{\eta}_y(t)\nonumber\\
\frac{\partial \theta}{\partial t}=& \tilde{\eta}_\theta,	
\end{align}
where the correlation of the thermal noise follows \cref{eq:fdt}.
\subsection{Purturbative Expansion}
\label{ssec:purturbative_expansion}

Defining the vector $\bfR\equiv (x,y)^T$, the equation takes the
simple form
\begin{equation}
  \label{eq:langevin_harmonic_vector}
  \overset{\bm .}{\bfR}=-\k \left[\barG \id + \frac{\D \G}{2} \doverline{\matR}(t)
  \right] \bfR(t) +\bfet(t)
\end{equation}

To solve the above equation, we use the perturbative expansion 
\begin{equation}
  \label{eq:perturbation_series}
  \bfR(t)=\bfR_0(t)- \left(\frac{\k \D \G}{2}\right) \bfR_1(t)
  +\left(\frac{\k \D \G}{2}\right)^2 \bfR_2(t)+\order\left(\frac{\k \D \G}{2}\right)^3
\end{equation}
Substituting \cref{eq:perturbation_series} in
\cref{eq:langevin_harmonic_vector} and keeping up to the linear order
in $\k \D\G/2$ we obtain the equations for
$\bfR(t)$ and $\bfR_1(t)$ as
\begin{equation}
  \label{eq:solution_series}
\begin{split}
  \overset{\bm .}{\bfR}_0&=-\k \barG \bfR_0(t) +\bfet(t)\\
  \overset{\bm .}{\bfR}_1&=-\k \barG \bfR_1(t)
  +\doverline{\matR}(t)\bfR_0(t)\\
  \overset{\bm .}{\bfR}_2&=-\k \barG \bfR_2(t)
  +\doverline{\matR}(t)\bfR_1(t)\\
\end{split}
\end{equation}

The solutions for the \cref{eq:solution_series} together with the
initial condition $\bfR(0)=0$ take the form
\begin{equation}
  \label{eq:solution_series_1}
\begin{split}
  \bfR_0(t)&=\int_0^t \diff t' e^{-\k \barG (t-t')}\bfet(t)\\
  \bfR_1(t)&=\int_0^t \diff t' e^{-\k \barG
    (t-t')}\;\doverline{\matR}(t)\;\bfR_0(t)\\
  \bfR_2(t)&=\int_0^t \diff t' e^{-\k \barG (t-t')}\;\doverline{\matR}(t)\;\bfR_1(t)
\end{split}
\end{equation}

In explicit form, the equal time correlation matrix $R_i(t) R_j(t)$ is then given by 
\begin{equation}
  \label{eq:msd_gen_exp}
\begin{split}
\la R_i(t) R_j(t) \ra_{\bfet,\theta}= \la R_{0,i}(t)
R_{0,j}(t)\ra_{\bfet,\theta} -\left(\frac{\k \D \G}{2}\right) \la
R_{0,i}(t)R_{1,j}(t)\ra_{\bfet,\theta}\\
+\left(\frac{\k \D
    \G}{2}\right)^2 \left[\left\la R_{1,i}(t)R_{1,j}(t)\right\ra_{\bfet,\theta}+2
\left\la R_{0,i}(t)R_{2,j}(t)\right\ra_{\bfet,\theta}\right] +\order\left(\frac{\k \D \G}{2}\right)^3
\end{split}
\end{equation}
where we have used the fact that
$\la R_{0,i}R_{1,j} \ra=\la R_{0,j}R_{1,i}\ra$. Further, note that the
thermal noise correlation given in \cref{eq:fdt} gives an additional
factor of $\k \D \G/2$ in the correlation terms
$\la R_{\a,i}(t) R_{\beta,j}(t)\ra $, where $\a,\beta$ denotes the
order of the perturbation series.

We next proceed to calculate this equal time correlation matrix using
the solutions in \cref{eq:solution_series_1}. The correlation matrix
of $\bfR_0(t)$ averaged over the translational and the rotational
noise is then given by
\begin{equation}
  \label{eq:correlation_bfR}
\begin{split}
  \la \bfR_0(t) \bfR_0(t) \ra_{\bfet,\theta} =\int_0^t \diff t' \int_0^t \diff t''
  e^{-\k \barG (t-t')} e^{-\k \barG (t-t'')} \la \bfet(t') \bfet(t'')
  \ra_{\bfet,\theta},
\end{split}
\end{equation}
where in correlation of the thermal noise is understood as an outer
product of the variable $\eta_x$ and $\eta_y$.
Using \cref{eq:fdt}, the calculation is straight forward and the final
form of the correlation matrix is given by

\begin{equation}
  \label{eq:correlation_bfR_final}
\begin{split}
  \la \bfR_0(t) \bfR_0(t) \ra_{\bfet,\theta} &= \frac{\kb T}{\k} \id
  \left(1- e^{-2 \k \barG t} \right) \\
&+ \D D \;\doverline{\matR}(\theta_0) 
\left(\frac{e^{-4 D_\theta t}-e^{-2 \k \barG t}}{2\k \barG-4 D_\theta}\right)
\end{split}
\end{equation}
More explicitly, the mean-square displacement along the $x$ and $y$
direction are given by
\begin{equation}
  \label{eq:msd_x_y}
\begin{split}
  \la x^2_0(t)\ra_{\bfet,\theta} = \frac{\kb T}{\k} 
  \left(1- e^{-2 \k \barG t} \right) 
+ \D D \cos 2\theta_0
\left(\frac{e^{-4 D_\theta t}-e^{-2 \k \barG t}}{2\k \barG-4
    D_\theta}\right) \\
\intertext{and}
  \la y^2_0(t)\ra_{\bfet,\theta} = \frac{\kb T}{\k} 
  \left(1- e^{-2 \k \barG t} \right) 
- \D D \cos 2\theta_0
\left(\frac{e^{-4 D_\theta t}-e^{-2 \k \barG t}}{2\k \barG-4
    D_\theta}\right) \\
\end{split}
\end{equation}
The cross-correlation function $x_0(t)y_0(t)$ reads
\begin{equation}
  \label{eq:corr_xy}
\begin{split}
  \la x_0(t)y_0(t)\ra_{\bfet,\theta} = \D D \sin 2\theta_0
\left(\frac{e^{-4 D_\theta t}-e^{-2 \k \barG t}}{2\k \barG-4
    D_\theta}\right) 
\end{split}
\end{equation}
\begin{figure}[!t]
  \centering
    \includegraphics[width=0.8\linewidth]{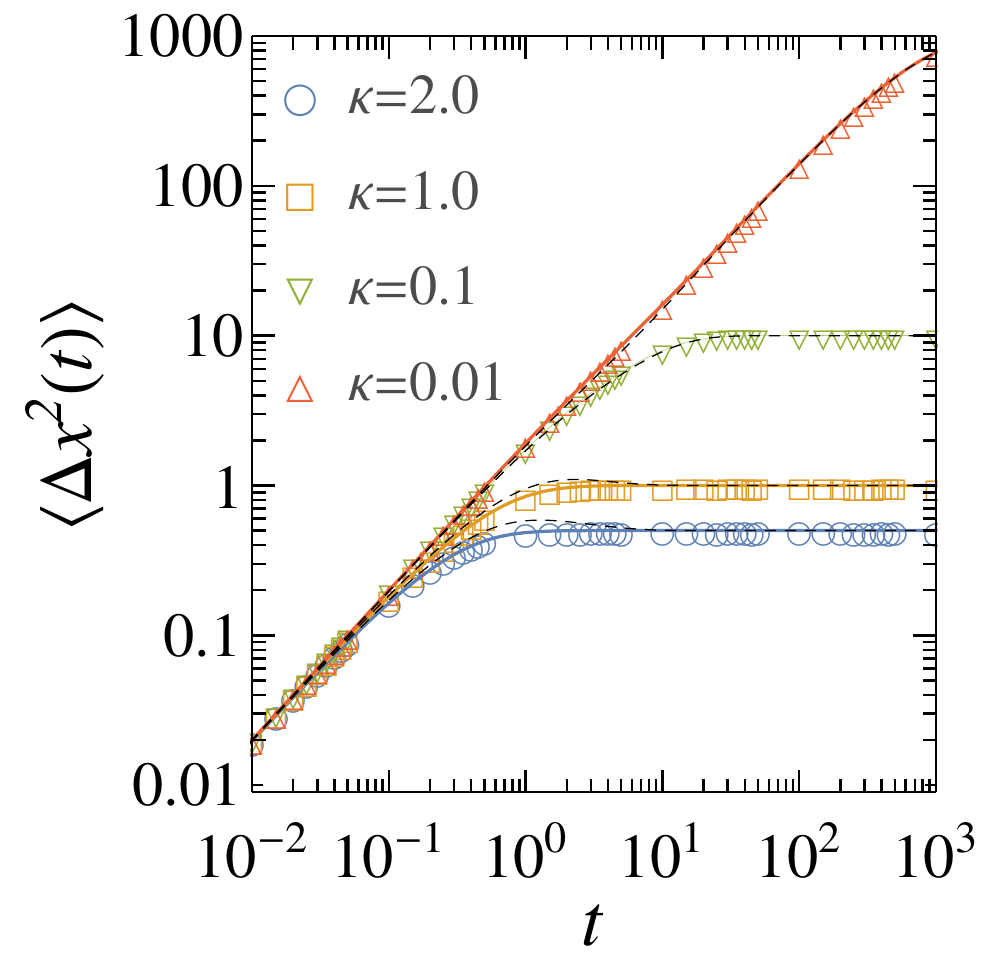}
    \caption{Plot of the mean-square displacement along the $x$
      direction of harmonically trapped anisotropic particle for
      different choices of the stiffness of the harmonic potential as
      indicated in the legend. The translational diffusivities and the
      rotational diffusion constant were kept fixed at
      $D_\parallel=1$, $D_\bot=0.5$ and $D_\theta=0.1$ in all the
      cases. The initial orientation of the particles were also fixed
      at $\theta_0=0$. The solid lines are plots of
      \cref{eq:msd_x_final} and the dashed lines are plots of
      \cref{eq:msd_x_approx} with the appropriate values of
      $\kappa,D_\parallel,D_\bot$ and $D_\theta$.}
  \label{fig:msd_x_y_harmonic}
\end{figure}

In the limit of $\k \to 0$, \cref{eq:msd_x_y,eq:corr_xy} reproduces
the correct result of a free diffusion of an anisotropic particle
given in \cref{eq:xsq_free_1,eq:ysq_free_1}. On the other hand, for
$\D \G \to 0$ \cref{eq:msd_x_y,eq:corr_xy} yeilds the correlation
matrix for an isotropic Brownian particle in a harmonic trap.

Our next attempt is to look into the correction to the above expression
that comes from $\bfR_1(t)$ and $\bfR_2(t)$. For this, we rewrite the
solutions for $\bfR_1(t)$ and $\bfR_2(t)$ in explicit form as
\begin{equation}
  \label{eq:solution_R1_explicit}
\begin{split}
  R_{1,i}(t)=\int_0^t \diff t' e^{-\k \barG (t-t')} \sum_j
  \matR_{ij}(t')R_{0,j}(t')\\
R_{2,i}(t)=\int_0^t \diff t' e^{-\k \barG (t-t')} \sum_j
\matR_{ij}(t')R_{1,j}(t')
\end{split}
\end{equation}
where the subscripts are for the two spatial dimensions and can take
the values $1$ and $2$. Using \cref{eq:msd_gen_exp}, we proceed to
calculate the terms $\la R_{0,i}(t)
R_{0,j}(t)\ra_{\bfet,\theta}$,$\la
R_{1,i}(t) R_{1,j}(t)\ra_{\bfet,\theta}$
and $\la R_{2,i}(t) R_{2,j}(t)\ra_{\bfet,\theta}$.  The detailed
calculation of the three terms are presented in the
\cref{appendix:appendix_1,appendix:appendix_2,appendix:appendix_3,appendix:appendix_4},respectively. In
deriving the results presented in the appendices, we have utilized the
more general form of the identity given in \cref{eq:identity_1}:
\begin{equation}
\label{eq:identity_2}
\la e^{i m \D \h(t')-i n \D
  \h(t'')}\ra_{\theta}=e^{-D_\h (m^2 t'+n^2 t''-2 m n \min (t',t''))}
\end{equation}
Using the above relation, the averages of the
trigonometric functions over the rotational noise take the form
\begin{equation}
\label{eq:identity_2}
\begin{split}
\la \cos 2[\h(t') -\h(t'')]\ra_\h&=e^{-4 D_\h(t'+t''-2 \min(t',t'') }\\
\la \cos 2[\h(t') +\h(t'')]\ra_\h&=\cos 4 \h_0 e^{-4 D_\h(t'+t''+2
  \min(t',t'') }\\
\la \sin 2[\h(t') +\h(t'')]\ra_\h&=\sin 4 \h_0 e^{-4 D_\h(t'+t''+2
  \min(t',t'') }\\
\la \sin 2[\h(t') -\h(t'')]\ra_\h&=0\\
\end{split}
\end{equation}

\begin{figure}[!t]
  \centering
  \includegraphics[width=0.8\linewidth]{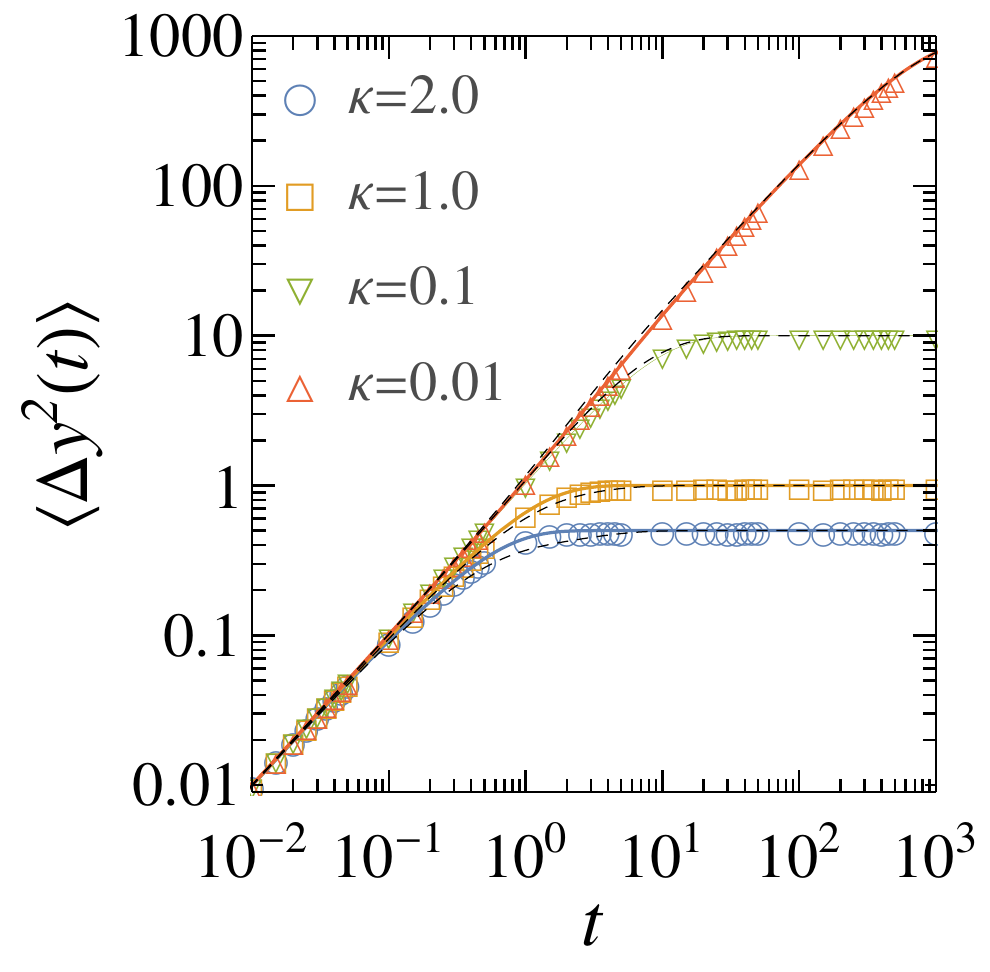}
 \caption{Plot of the mean-square displacement along the $y$ 
    direction of harmonically trapped anisotropic particle for
    different choices of the stiffness of the harmonic potential as
    indicated in the legend. The translational
    diffusivities and the rotational diffusion constant were kept
    fixed at $D_\parallel=1$, $D_\bot=0.5$ and $D_\theta=0.1$ in all
    the cases. The initial orientation of the particles were also fixed
      at $\theta_0=0$.The solid lines are plots of \cref{eq:msd_y_final} and
    the dashed lines are plots of \cref{eq:msd_y_approx} with the
    appropriate values of $\kappa,D_\parallel,D_\bot$ and $D_\theta$.}
  \label{fig:msd_y_harmonic}
\end{figure}
The final form of the expressions is given by
\begin{equation}
  \label{eq:x0_x1}
\begin{split}
 & \la x_0(t) x_1(t) \ra_{\bfet,\h}=\la y_0(t) y_1(t)
  \ra_{\bfet,\h}=\\
&\left(\frac{\kb T}{\k}\right) \cos 2\h_0 \;
 \left(\frac{e^{-4 D_\h
      t}-e^{-2\k \barG t}}{(2\k \barG-4 D_\h)} -\frac{e^{-2\k \barG
      t}-e^{-(2 \k \barG+4 D_\h) t}}{4 D_\h} \right)\\
&+2\left(\frac{\kb T}{\k}\right)\;\left(\frac{\k \D \G}{2}\right)
\left(\frac{1-e^{-2 \k \barG t}}{2 \k \barG (2 \k \barG +4
  D_\h)}-\frac{e^{-2 \k \barG t}-e^{-(2 \k \barG+4 D_\h) t}}{4 D_\h (2
  \k \barG +4 D_\h)}\right)\\
\end{split}
\end{equation}
\begin{equation}
\label{eq:x1_sq_final}
\begin{split}
  \la x^2_1(t) \ra_{\bfet,\theta}&=\la y^2_1(t) \ra_{\bfet,\theta}=\left( \frac{ \kb T}{\k} \right)
  \left[ \frac{1-e^{-2 \k \barG t}}{\k \barG (2 \k \barG +4
      D_\h)}-\frac{t e^{-2 \k \barG t}}{4D_\h} \right. \\
&\quad \quad \qquad \qquad \qquad \left. +\k \barG\frac{e^{-2 \k
        \barG t}-e^{-(2 \k \barG + 4 D_\h) t}}{4 D^2_\h (2 \k \barG +4 D_\h)} \right]
\end{split}
\end{equation}

\begin{equation}
  \label{eq:x0_x2}
  \begin{split}
 \left\la x_0(t) x_2(t) \right\ra_{\bfet,\theta}&=\left\la y_0(t) y_2(t) \right\ra_{\bfet,\theta}
=\left(\frac{\kb
     T}{\k}\right)\left[\frac{1-e^{-2\k
      \barG t}}{2 \k \barG (2\k\barG +4 D_\h)}\right.\\
&\left.-\frac{te^{-2\k \barG t}
  }{4 D_\h}  +\frac{2\k \barG}{4 D_\h}\left(1-e^{-4 D_\h
      t}\right)  \right] \\  
  \end{split}
\end{equation}

In the limit of $\k \to 0$, both
$\la y_1^2(t) \ra =\la x_1^2(t) \ra=0$.

The final expression for the mean-square displacement along the $x$
is given by
\begin{equation}
  \label{eq:msd_x_final}
  \begin{split}
    \la x^2(t) \ra_{\bfet,\h}&=\left(\frac{\kb
        T}{\k}\right)\left[\left(1-e^{-2 \k \barG t}\right) +\left(\frac{\k
        \D \G}{2}\right) \cos 2\h_0 \right.\\
      &\left( \frac{e^{-4 D_\h t}-e^{-2
          \k \barG t}}{2 \k \barG - 4 D_\h}+\frac{e^{-2 \k \barG
          t}-e^{-(2 \k \barG +4 D_\h) t}}{4 D_\h}\right)\\
     &+\left(\frac{\k
        \D \G}{2}\right)^2 \left( \frac{1}{4 D_\h^2} e^{-2\k\barG t}\left( 1-e^{-4 D_\h t} \right) -\frac{ t}{ D_\h} e^{-2\k
        \barG t}  \right) \\
     & \left. +\order\left(\frac{\k
        \D \G}{2}\right)^3 \right]
  \end{split}
\end{equation}

and that along the $y$-direction is given by 
\begin{equation}
  \label{eq:msd_y_final}
  \begin{split}
    \la y^2(t) \ra_{\bfet,\h}&=\left(\frac{\kb
        T}{\k}\right)\left[\left(1-e^{-2 \k \barG t}\right) -\left(\frac{\k
        \D \G}{2}\right) \cos 2\h_0 \right.\\
      &  \left( \frac{e^{-4 D_\h t}-e^{-2
          \k \barG t}}{2 \k \barG - 4 D_\h}+\frac{e^{-2 \k \barG t}-e^{-(2 \k \barG +4 D_\h) t}}{4 D_\h}\right)\\
    &+\left(\frac{\k
        \D \G}{2}\right)^2 \left(\frac{1}{4 D_\h^2} e^{-2\k\barG t}\left( 1-e^{-4 D_\h t} \right) -\frac{ t}{ D_\h} e^{-2\k
        \barG t}  \right) \\
     &\left.+\order\left(\frac{\k \D \G}{2}\right)^3 \right]
  \end{split}
\end{equation}

\subsection{Mean-square displacement for large rotational diffusion constant}

In this section we present an alternate expression for mean-square
displacement of an anisotropic particle which is valid for whose
rotational diffusion constant is large as compared to the inverse
times scales $\k \barG $ and $\k \D\G$. In such a scenario, since the
particle rotates faster, the mobility of the anisotropic particle is
an average mobility over the rotational noise. We start our analysis
with \cref{eq:solution_langevin_R}, but we set $\bfR(0)=0$. To proceed
further, and in particular to look at the asymptotic limit of the
correlations, we define the variable $u=(t-t')/t$. In terms of the new
variable $u$, the solution for $\bfR(t)$ takes the form
\begin{equation}
  \label{eq:solution_langevin_R}
  \bfR(t)=t\int_0^1 \diff u\, e^{-\k \bar{\Gamma}\,\id\, t\, u}
  e^{-\frac{\k}{2} \D \Gamma\int_{t(1-u)}^t \,\diff t'' \,
   \doverline{\matR}[\theta(t'')] } \,\bfet[t(1-u)]
\end{equation}
The equal-time correlation is then given by
\begin{equation}
  \label{eq:correlation_approx}
\begin{split}
  \la \bfR(t) \bfR(t) \ra_{\bfet}&= \int_0^1 \diff u \int_0^1 \diff u' e^{-\k
    \bar{\Gamma}\,\id\, t\, u} e^{-\k \bar{\Gamma}\,\id\, t\, u'} e^{-\frac{\k}{2} \D \Gamma\int_{t(1-u)}^t \,\diff t'' \,
   \doverline{\matR}[\theta(t'')] } \\
&e^{-\frac{\k}{2} \D \Gamma\int_{t(1-u')}^t \,\diff t'' \,
   \doverline{\matR}[\theta(t'')] } \,\,\la
 \bfet[t(1-u)]\bfet[t(1-u')]\ra_{\bfet}
\end{split}
\end{equation}
The correlation of the thermal noise in the transformed variable is 
\begin{equation}
  \label{eq:noise_correlation_u}
  \la \bfet[t(1-u)] \bfet[t(1-u')] \ra_{\bfet}= \frac{2\kb T
  }{t}\doverline{\bgma}[t(1-u)] \delta(u-u')
\end{equation}
Substituting the noise correlation into \cref{eq:correlation_approx}
and integration over $u'$ we get
\begin{equation}
  \label{eq:correlation_approx}
 \begin{split}
  \la \bfR(t) \bfR(t) \ra &= 2 \kb T t \int_0^1 \diff u e^{- 2 \k \bar{\Gamma}\,\id\, t\, u} e^{-\k \D \Gamma\int_{t(1-u)}^t \,\diff t'' \,
   \doverline{\matR}[\theta(t'')] } \times\\
    &\phantom{16pt}\;\;\,\left[ \barG \id + \frac{\D \G }{2} \doverline{\matR}[\theta
      (t(1-u))]\right]
\end{split}
\end{equation}
In the asymptotic limit, the integral is dominated by small values of
$u$, the integral in the exponential from $t(1-u)$ to $t$ is
vanishingly small and can be set to zero. Further, we set
$\doverline{\matR}[\theta(t(1-u))] \approx
\doverline{\matR}[\theta(t)]$. Consequently, the correlation matrix
averaged over the translational noise take the form
\begin{equation}
  \label{eq:correlation_approx_1}
 \begin{split}
  \la \bfR(t) \bfR(t) \ra = 2 \kb T t \int_0^1 \diff u e^{- 2 \k \bar{\Gamma}\,\id\, t\, u}  
    \left[ \barG \id + \frac{\D \G }{2} \doverline{\matR}[\theta
      (t)]\right]
\end{split}
\end{equation}
and performing the average over the rotational noise and the integral
over $u$ we arrive at
\begin{equation}
  \label{eq:correlation_approx_1}
 \begin{split}
  \la \bfR(t) \bfR(t) \ra_{\bfet,\theta} = 2 \kb T t
  \left(\frac{1-e^{-2 \k \barG t}}{2 \k \barG t}\right)
    \left( \barG \id + \frac{\D \G }{2} \la \doverline{\matR}[\theta(t)]\ra_{\theta}\right)
\end{split}
\end{equation}
Simplifying the result and using \eqref{eq:identity_1} we arrive at 
\begin{equation}
  \label{eq:correlation_approx_2}
 \begin{split}
  \la \bfR(t) \bfR(t) \ra_{\bfet,\theta} =  \frac{\kb T }{\k \barG}
  \left(1-e^{-2 \k \barG t}\right)
    \left( \barG \id + \frac{\D \G }{2} \, \doverline{\matR}(\theta_0) \, e^{-4 D_\theta t}\right).
\end{split}
\end{equation}
The mean-square displacement in the explicit form is given by
\begin{equation}
  \label{eq:msd_x_approx}
   \la \D x^2(t) \ra_{\bfet,\theta} =  \frac{\kb T }{\k}
  \left(1-e^{-2 \k \barG t}\right)
    \left( 1  + \frac{\D \G }{2\barG} \, \cos 2\theta_0 \, e^{-4 D_\theta t}\right).
\end{equation}

\begin{equation}
  \label{eq:msd_y_approx}
   \la \D y^2(t) \ra_{\bfet,\theta} =  \frac{\kb T }{\k}
  \left(1-e^{-2 \k \barG t}\right)
    \left( 1  - \frac{\D \G }{2\barG} \, \cos 2\theta_0 \, e^{-4 D_\theta t}\right).
\end{equation}
and 
\begin{equation}
  \label{eq:msd_xy_approx}
   \la \D x(t) \D y(t) \ra_{\bfet,\theta} =  \frac{\kb T }{\k}
  \left(1-e^{-2 \k \barG t}\right)
    \left( \frac{\D \G }{2\barG} \, \sin 2\theta_0 \, e^{-4 D_\theta t}\right).
\end{equation}
Note that there is striking difference between the
\cref{eq:msd_x_approx,eq:msd_y_approx} and that of \cref{eq:msd_x_final,eq:msd_y_final}
with respect to the limit of $\kappa \to 0$. While the later
expressions correctly reproduces the free diffusion of the anisotropic
particle, the limit of $\k \to 0$ in \cref{eq:correlation_approx_2}
yields the correct asymptotic result by setting $e^{-4 D_\h t} \to 0$:
\begin{equation}
  \label{eq:asymptotic_approx}
  \la \D x^2(t)\ra=\la \D y^2(t) \ra =2 \kb T \barG t  \; \;\; \textrm{and}
  \; \; \;\la \D x(t) \D y(t) \ra_{\bfet,\theta} =0.
\end{equation}
 \subsection{Persistence Probability}
\label{ssec:persistence_harmonic}
We now turn our attention to the persistence probability of the
harmonically trapped ellipsoidal particle. For this, we focus on the
two time correlation function $\la x(t_1)x(t_2)
\ra_{\bfet,\theta}$. Using the perturbation series given in
\cref{eq:perturbation_series} we have up to order $\order(\k\D\G/2)$
\begin{equation}
  \label{eq:x_two_time_correlation}
\begin{split}
  \la x(t_1)x(t_2)\ra_{\bfet,\theta}&= \la
  x_0(t_1)x_0(t_2)\ra_{\bfet,\theta}\\
&-\left(\frac{\k \D\G}{2}\right)\left[
    \la x_0(t_1)x_1(t_2)\ra_{\bfet,\theta}+ \la
    x_0(t_2)x_1(t_1)\ra_{\bfet,\theta} \right]
\end{split}
\end{equation}
where $t_1>t_2$.  The correlation functions
$ \la x_0(t_1)x_1(t_2)\ra_{\bfet,\theta}$ and
$ \la x_0(t_2)x_1(t_1)\ra_{\bfet,\theta}$ are equal only in the
asymptotic limit, that is for $t_1$ and $t_2 $ large. In this limit,
the expression for the two time correlation function takes the form
\begin{equation}
  \label{eq:x_two_time_correlation_1}
\begin{split}
  \la x(t_1)x(t_2)\ra_{\bfet,\theta}&= \la
  x_0(t_1)x_0(t_2)\ra_{\bfet,\theta}\\
&-\left(\k \D\G\right)\left[
    \la x_0(t_1)x_1(t_2)\ra_{\bfet,\theta}\right]
\end{split} 
\end{equation}
The correlation functions $\la x_0(t_1) x_0(t_2) \ra_{\bfet,\theta}$
and $\la x_0(t_1)x_1(t_2)\ra_{\bfet,\theta}$ are derived in 
in \cref{appendix:appendix_1} (see  \cref{eq:eq_1} ) and
in  \cref{appendix:appendix_4} (see \cref{eq:eq_10}),
respectively. For completeness, we
quote the main results here. 
\begin{equation}
  \label{eq:x0_t1_x0_t2}
  \begin{split}
      \left\la x_0(t_1) x_0(t_2) \right\ra_{\bfet,\theta}&=\frac{\kb
    T}{\k}  \left[e^{-\k \barG |t_1-t_2|} -e^{-\k \barG
      (t_1+t_2)} \right]\\
& +\left(\frac{\kb T}{\k}\right)\;\k\D \G \cos 2\h_0 e^{-\k \barG t_1} \left[\frac{e^{(\k \barG
      -4 D_\h)t_2}-e^{-\k \barG t_2}}{2\k \barG -4 D_\h}\right]
  \end{split}
\end{equation}
\begin{equation}
\label{eq:x0_t1_x1_t2}
\begin{split}
\la x_0(t_1) x_1(t_2) \ra_{\bfet,\theta}&= \left(\frac{\kb
     T}{\k}\right) \cos 2\h_0 \; e^{-\k \barG t_1} \;\left(\frac{e^{(\k
       \barG -4D_\h )t_2} -e^{-\k \barG t_2}}{2 \k \barG - 4 D_\h}\right.\\
&\left.   -\frac{e^{-\k \barG t_2}-e^{-(4 D_\h+\k \barG) t_2}}{4 D_\h}
 \right) +\left(\frac{\kb T}{\k}\right)\;\left(\frac{\D \G}{2\barG}\right)
 e^{-\k \barG t_1}\\
&\left[\frac{e^{\k \barG t_2}-e^{-\k \barG t_2}}{(2\k \barG +4 D_\h)}
    -\frac{2\k\barG}{4D_\h}\frac{e^{-\k \barG t_2}-e^{-(2 \k \barG +4
        D_\h) t_2}}
{(\k \barG +4 D_\h)}\right]\\
\end{split}
\end{equation}

\begin{figure}[!t]
  \centering
  \includegraphics[width=\linewidth]{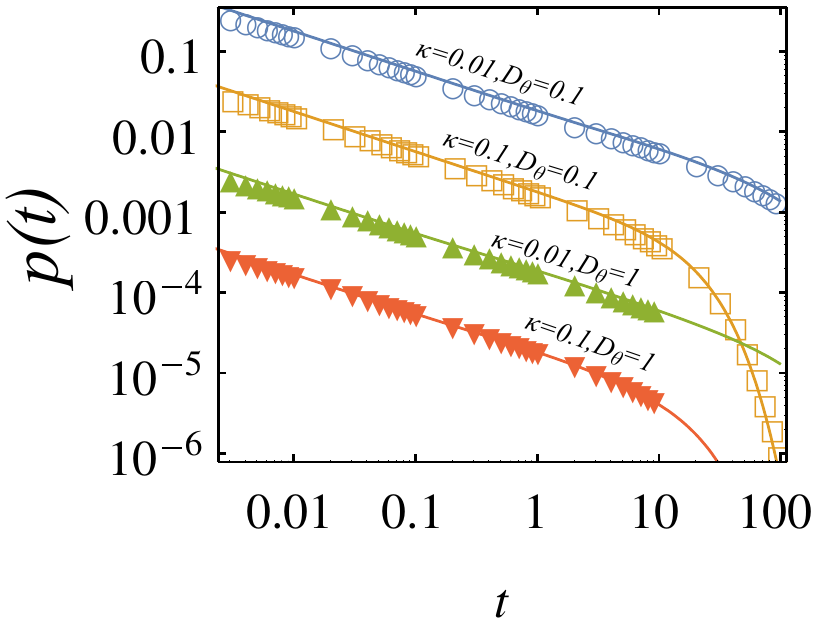}
  \caption{Plot of the survival probability $p(t)$ of a harmonically
    trapped anisotropic particle for different choices of the
    rotational diffusion constant and the stiffness of the potential,
    as indicated alongside each plot. The solid lines are plots of
    \cref{eq:persistence_probability} with the appropriate values of
    $\kappa,D_\parallel,D_\bot$ and $D_\theta$. While the rotational
    diffusion constant and the spring stiffness was varied, the
    translational diffusivities  and the initial angle $\theta_0$ were
    fixed at values $D_\parallel=1$, $D_\bot=0.5$ and $\theta_0=0$. }
  \label{fig:persistence_harmonic}
\end{figure}

Note that in calculating the two time correlation function up to an
order $\order(\k \D \G)$, we will use only the first term appearing in
\cref{eq:x0_t1_x1_t2}. Looking at \cref{eq:x_two_time_correlation_1}
and \cref{eq:x0_t1_x0_t2,eq:x0_t1_x1_t2} it is clear that the first
term contained in the parenthesis in \cref{eq:x0_t1_x1_t2} cancels
with the term proportional to $\k \D\ G$ in \cref{eq:x0_t1_x0_t2}. The
final expression for $\la x(t_1)x(t_2)\ra_{\bfet,\theta}$ reads
\begin{equation}
  \label{eq:x_two_time_correlation_final}
\begin{split}
  \la x(t_1)x(t_2)\ra_{\bfet,\theta}&= \left( \frac{2\kb
    T}{\k}\right) e^{-\k \barG t_1}  \bigg[\sinh \k \barG t_2 \\
&+\left( \frac{\k \D \G}{2}\right) \cos 2\h_0 \; e^{-\k \barG t_2}\left(
   \frac{1-e^{-4 D_\h t_2}}{4 D_\h}  \right) \bigg]
\end{split} 
\end{equation}

As before, defining the variable $X(t)=x(t)/\sqrt{\la x^2
  \ra}_{\bfet,\theta}$, the correlation function of $\la X(t_1) X(t_2)
\ra_{\bfet,\theta}$ is given by
\begin{equation}
  \label{eq:X_t1_X_t2}
\begin{split}
  \la X(t_1) X(t_2) \ra_{\bfet,\theta}&=\\
&\mkern-54mu \frac{e^{-\k \barG
      t_1/2}}{e^{-\k \barG t_2/2}}\left[\frac{\sinh \k \barG t_2 
+\left( \frac{\k \D \G}{2}\right) \cos 2\h_0 \; e^{-\k \barG t_2}\left(
   \frac{1-e^{-4 D_\h t_2}}{4 D_\h}  \right)}{\sinh \k \barG t_2 
+\left( \frac{\k \D \G}{2}\right) \cos 2\h_0 \; e^{-\k \barG t_2}\left(
   \frac{1-e^{-4 D_\h t_2}}{4 D_\h}  \right)}\right]^{1/2}
\end{split}
\end{equation}
Using the transformation
$e^T=e^{\k \barG t} \left[\sinh \k \barG t +\left( \frac{\k \D
      \G}{2}\right) \cos 2\h_0 \; e^{-\k \barG t}\left(
    \frac{1-e^{-4 D_\h t}}{4 D_\h} \right) \right]^{1/2}$
for an imaginary time variable $T$, the correlation function $\la
X(T_1) X(T_2) \ra_{\bfet,\theta}$ becomes a stationary correlator : $\la
X(T_1) X(T_2) \ra_{\bfet,\theta} =e^{-(T_1-T_2)/2}$ and the
corresponding persistence probability is given by 
\begin{equation}
  \label{eq:persistence_probability}
  p(t) \sim  \frac{\sqrt{\k}e^{-\k \barG t/2}}{\left[\sinh \k \barG t +\left( \frac{\k \D
      \G}{2}\right) \cos 2\h_0 \; e^{-\k \barG t}\left(
    \frac{1-e^{-4 D_\h t}}{4 D_\h} \right) \right]^{1/2}}
\end{equation}
In the limit of $\D\G \to 0$ the equation correctly reproduces the
persistence of probability of an isotropic particle in the presence of
a harmonic trap. \cite{chakraborty2007} The other limit of $\k \to 0$
reproduces the persistence probability of a free anisotropic particle
derived in \cref{eq:per_prob_real_time}.

\section{Conclusion}
\label{sec:conclusion}

In summary, we have determined the persistence probability of an
anisotropic particle in two spatial dimensions, in the presence as
well as in the absence of a confining harmonic potential. The two time
correlation functions of the position of the particle has been
calculated in both cases. In the case of a harmonically confined
particle, a purtubative solution has been provided for the
correlation functions. The persistence probability is computed from
the two-time correlation function using suitable transformations in
space and time. The determination of the rotational and the
translational diffusion coefficients have been explicitly carried out
for an anisotropic particle that undergoes free Brownian
motion. Additionally, the analytical results have been confirmed by
numerical simulation of the underlying stochastic dynamics. 
\bibliography{references_persistence}

\cleardoublepage

\appendixpage
\onecolumngrid
\begin{appendices}
\numberwithin{equation}{section}
\section{Calculation of $\mathbf{\la R_{0,i}(t) R_{0,j}(t) \ra}$.}
\label{appendix:appendix_1}

\begin{equation}
  \label{eq:correlation_bfR_1}
\begin{split}
  \la \bfR_0(t) \bfR_0(t) \ra_{\bfet,\theta} =\int_0^t \diff t' \int_0^t \diff t''
  e^{-\k \barG (t-t')} e^{-\k \barG (t-t'')} 
\left[ \barG \id +
    \frac{\D \G}{2} \doverline{\matR}(t') \right] \delta (t'-t'')
\end{split}
\end{equation}

\begin{equation}
  \label{eq:correlation_bfR_2}
  \la \bfR_0(t) \bfR_0(t) \ra_{\bfet,\theta} =2 \kb T e^{-2 \k \barG
    t} \int_0^t \diff t' e^{2\k \barG t'}  \left[ \barG \id +
    \frac{\D \G}{2} \left\la \doverline{\matR}(t')\right\ra_{\bfet,\theta} \right]
\end{equation}

\begin{equation}
  \label{eq:correlation_bfR_3}
\begin{split}
  \la \bfR_0(t) \bfR_0(t) \ra_{\bfet,\theta} = \frac{\kb T}{\k} \id
  \left(1- e^{-2 \k \barG t} \right) 
+ 2 \kb T e^{-2 \k \barG
    t} \int_0^t \diff t' e^{2\k \barG
    t'} \;\frac{\D \G}{2} \;\doverline{\matR}(\theta_0) \;e^{-4
    D_\theta t'}
\end{split}
\end{equation}

\begin{equation}
  \label{eq:correlation_bfR_4}
\begin{split}
  \la \bfR_0(t) \bfR_0(t) \ra_{\bfet,\theta} = \frac{\kb T}{\k} \id
  \left(1- e^{-2 \k \barG t} \right) 
+ \D D \;\doverline{\matR}(\theta_0) \;e^{-2 \k \barG
    t} \left(\frac{e^{(2\k \barG-4 D_\theta) t}-1}{2\k \barG-4 D_\theta}\right)
\end{split}
\end{equation}

\begin{equation}
  \label{eq:correlation_bfR_4}
\begin{split}
  \la \bfR_0(t) \bfR_0(t) \ra_{\bfet,\theta} = \frac{\kb T}{\k} \id
  \left(1- e^{-2 \k \barG t} \right) 
+ \D D \;\doverline{\matR}(\theta_0) 
\left(\frac{e^{-4 D_\theta t}-e^{-2 \k \barG t}}{2\k \barG-4 D_\theta}\right)
\end{split}
\end{equation}

\section{Calculation of $\la R_{0,i}(t) R_{1,j}(t) \ra$}
\label{appendix:appendix_2}

Calculation of $\la R_{0,i}(t) R_{1,j}(t) \ra$.

\begin{equation}
  \label{eq:eq_1}
\begin{split}
  \left\la R_{0,i}(t_1) R_{0,j}(t_2) \right\ra_{\bfet}&=\int_0^{t_1}
  \diff t'_1 \int_0^{t_2} \diff t'_2\; e^{-\k \barG (t_1-t'_1)} \;
  e^{-\k \barG (t_2-t'_2)} \;
  \la \eta_i(t'_1)\eta_j(t'_2) \ra_{\bfet}\\
  \left\la R_{0,i}(t_1) R_{0,j}(t_2) \right\ra_{\bfet}&=2 \kb T e^{-\k
    \barG (t_1+t_2)} \int_0^{t_1} \diff t'_1 \int_0^{t_2} \diff t'_2\;
  e^{\k \barG (t'_1+t'_2)} \;
  \left[\barG \delta_{ij} +\frac{\D \G}{2} \matR_{ij}(t'_1)\right] \;\delta(t'_1-t'_2)\\
  \left\la R_{0,i}(t_1) R_{0,j}(t_2) \right\ra_{\bfet}&=2 \kb T \barG
  \delta_{ij} e^{-\k \barG (t_1+t_2)} \int_0^{t_1} \diff t'_1
  \int_0^{t_2} \diff t'_2\; e^{\k \barG (t'_1+t'_2)}
  \;\delta(t'_1-t'_2)\; \\
  &+2 \kb T\;\frac{\D \G}{2} e^{-\k \barG (t_1+t_2)} \int_0^{t_1}
  \diff t'_1 \int_0^{t_2} \diff t'_2\; e^{\k \barG
    (t'_1+t'_2)}  \matR_{ij}(t'_1) \;\delta(t'_1-t'_2)\\
  \left\la R_{0,i}(t_1) R_{0,j}(t_2) \right\ra_{\bfet}&=2 \kb T \barG
  \delta_{ij} e^{-\k \barG (t_1+t_2)} \int_0^{\min(t_1,t_2)} \diff
  t'_1 \;e^{2\k \barG t'_1} +2 \kb T\;\frac{\D \G}{2} e^{-\k \barG
    (t_1+t_2)} \int_0^{\min(t_1,t_2)} \diff t'_1 \;e^{2\k \barG t'_1}
  \matR_{ij}(t'_1) \\
  \left\la R_{0,i}(t_1) R_{0,j}(t_2) \right\ra_{\bfet}&=\frac{\kb
    T}{\k} \delta_{ij} \left[e^{-\k \barG |t_1-t_2|} -e^{-\k \barG
      (t_1+t_2)} \right] +\kb T\;\D \G e^{-\k \barG
    (t_1+t_2)}
  \int_0^{\min(t_1,t_2)} \diff t'_1  \;e^{2\k \barG t'_1}  \matR_{ij}(t'_1) \\
\end{split}
\end{equation}

\begin{equation}
  \label{eq:eq_2}
  \begin{split}
  \la R_{0,i}(t) R_{1,j}(t)\ra_{\bfet,\theta}&=\left \la R_{0,i}(t) \int_0^t \diff t'
    e^{-\k \barG (t-t')} \sum_k \matR_{jk}(t')R_{0,k}(t')
  \right\ra_{\bfet,\theta}\\
 \la R_{0,i}(t) R_{1,j}(t)\ra_{\bfet,\theta} &=\left \la  \int_0^t \diff t'
    e^{-\k \barG (t-t')} \sum_k \matR_{jk}(t')\left\la R_{0,i}(t)
    R_{0,k}(t') \right\ra_{\bfet}
  \right\ra_{\theta}
  \end{split}
\end{equation}
Using the final form of $\la R_{0,i}(t_1)R_{0,j}(t_2)\ra$ from
\cref{eq:eq_1} and identifying $t_1\equiv t$, $t' \equiv t_2$ with
$t'<t$ we get
\begin{equation}
  \label{eq:eq_3}
  \left\la R_{0,i}(t) R_{0,k}(t') \right\ra_{\bfet}=\frac{\kb
    T}{\k} \delta_{ik} \left[e^{-\k \barG (t-t')} -e^{-\k \barG
      (t+t')} \right] +\kb T\;\D \G e^{-\k \barG
    (t+t')}
  \int_0^{t'} \diff t'_1  \;e^{2\k \barG t'_1}  \matR_{ik}(t'_1) 
\end{equation}
Substituting \cref{eq:eq_3} in \cref{eq:eq_2} we get
\begin{equation}
  \label{eq:eq_4}
  \begin{split}
 \la R_{0,i}(t) R_{1,j}(t)\ra_{\bfet,\theta} &=\left \la  \int_0^t \diff t'
   e^{-\k \barG (t-t')} \sum_k \matR_{jk}(t')\left[
   \frac{\kb
    T}{\k} \delta_{ik} \left(e^{-\k \barG (t-t')} -e^{-\k \barG
      (t+t')} \right) +\kb T\;\D \G e^{-\k \barG
    (t+t')}
  \int_0^{t'} \diff t'_1  \;e^{2\k \barG t'_1}  \matR_{ik}(t'_1)\right]
\right\ra_{\theta}\\
\la R_{0,i}(t) R_{1,j}(t)\ra_{\bfet,\theta} &=\left(\frac{\kb T}{\k}\right) \matR_{ji}(\h_0) e^{-2 \k \barG t}\int_0^t \diff t'
   e^{-4 D_\h t'} \; \left(e^{2\k \barG t'} -1 \right) +\kb T\;\D \G e^{-2 \k \barG t}\int_0^t \diff t'
  \int_0^{t'} \diff t'_1  \;e^{2\k \barG t'_1}  \left \la \sum_k
    \matR_{jk}(t')\matR_{ik}(t'_1) \right\ra_{\theta}\\
  \la R_{0,i}(t) R_{1,j}(t)\ra_{\bfet,\theta} &=\left(\frac{\kb T}{\k}\right) \matR_{ji}(\h_0) e^{-2 \k \barG t}\int_0^t \diff t'
   \; \left(e^{(2\k \barG-4 D_\h) t'} -e^{-4 D_\h t'} \right) +\kb T\;\D \G e^{-2 \k \barG t}\int_0^t \diff t'
  \int_0^{t'} \diff t'_1  \;e^{2\k \barG t'_1}  \left \la \sum_k
    \matR_{jk}(t')\matR_{ik}(t'_1) \right\ra_{\theta}
  \end{split}
\end{equation}
For the mean-square displacement along the $x$ and the $y$ direction,
the second term in the last line of \cref{eq:eq_4} yeilds 
\begin{equation}
  \label{eq:eq_5}
\begin{split}
  \left \la \sum_k
    \matR_{ik}(t')\matR_{ik}(t'_1) \right\ra_{\theta}=&\la \cos 2\h(t')
  \cos 2\h(t'_1) +\sin 2\h (t') \sin 2 \h(t'_1) \ra_\h=\la \cos 2
  (\h(t')-\h(t'_1)) \ra_\h\\
  \left \la \sum_k
    \matR_{ik}(t')\matR_{ik}(t'_1) \right\ra_{\theta}&=e^{-4 D_\h (t'-t'_1)}
\end{split}
\end{equation}
On the other hand for $i \neq j$, the term
$ \left \la \sum_k \matR_{jk}(t')\matR_{ik}(t'_1)
\right\ra_{\theta}=0$.
Using \cref{eq:eq_5} the contribution to the mean-square displacement
along the $x$-direction becomes
\begin{equation}
  \label{eq:eq_6}
\begin{split}
  \la x_0(t) x_1(t)\ra_{\bfet,\theta}&=\left(\frac{\kb T}{\k}\right) \cos 2\h_0 e^{-2 \k \barG t}
   \; \left(\frac{e^{(2\k \barG-4 D_\h) t}-1}{(2\k \barG-4 D_\h)} -\frac{1-e^{-4 D_\h t}}{4 D_\h} \right) +\kb T\;\D \G e^{-2 \k \barG t}\int_0^t \diff t'
  \int_0^{t'} \diff t'_1  \;e^{2\k \barG t'_1}  e^{-4 D_\h (t'-t'_1)}\\
&=\left(\frac{\kb T}{\k}\right) \cos 2\h_0 \; \left(\frac{e^{-4 D_\h
      t}-e^{-2\k \barG t}}{(2\k \barG-4 D_\h)} -\frac{e^{-2\k \barG
      t}-e^{-(2 \k \barG+4 D_\h) t}}{4 D_\h} \right)+\kb T\;\D \G e^{-2 \k \barG t}\int_0^t \diff t'
  \;e^{-4 D_\h t'} \;\frac{e^{(2\k \barG +4 D_\h )t'}-1}{2\k \barG +4
    D_\h} \\
&=\left(\frac{\kb T}{\k}\right) \cos 2\h_0 \; \left(\frac{e^{-4 D_\h
      t}-e^{-2\k \barG t}}{(2\k \barG-4 D_\h)} -\frac{e^{-2\k \barG
      t}-e^{-(2 \k \barG+4 D_\h) t}}{4 D_\h} \right)+\kb T\;\D \G
e^{-2 \k \barG t} \left(\frac{e^{2 \k \barG t}-1}{2 \k \barG (2 \k \barG +4
  D_\h)}-\frac{1-e^{-4 D_\h t}}{4 D_\h (2 \k \barG +4 D_\h)}\right)\\
&=\left(\frac{\kb T}{\k}\right) \cos 2\h_0 \; \left(\frac{e^{-4 D_\h
      t}-e^{-2\k \barG t}}{(2\k \barG-4 D_\h)} -\frac{e^{-2\k \barG
      t}-e^{-(2 \k \barG+4 D_\h) t}}{4 D_\h} \right)+\kb T\;\D \G
\left(\frac{1-e^{-2 \k \barG t}}{2 \k \barG (2 \k \barG +4
  D_\h)}-\frac{e^{-2 \k \barG t}-e^{-(2 \k \barG+4 D_\h) t}}{4 D_\h (2 \k \barG +4 D_\h)}\right)\\
\end{split}
\end{equation}
and that along the $y$-direction takes the form
\begin{equation}
  \label{eq:eq_y0_y1}
  \begin{split}
    \la y_0(t)y_1(t) \ra_{\bfet,\theta}=-\left(\frac{\kb T}{\k}\right) \cos 2\h_0 \; \left(\frac{e^{-4 D_\h
      t}-e^{-2\k \barG t}}{(2\k \barG-4 D_\h)} -\frac{e^{-2\k \barG
      t}-e^{-(2 \k \barG+4 D_\h) t}}{4 D_\h} \right)+\kb T\;\D \G
\left(\frac{1-e^{-2 \k \barG t}}{2 \k \barG (2 \k \barG +4
  D_\h)}-\frac{e^{-2 \k \barG t}-e^{-(2 \k \barG+4 D_\h) t}}{4 D_\h (2 \k \barG +4 D_\h)}\right)\\
  \end{split}
\end{equation}
\section{Calculation of $\la R_{1,i}(t) R_{1,j}(t) \ra$}
\label{appendix:appendix_3}

The correlation matrix now takes the form
\begin{equation}
  \label{eq:correlation_matrix_R1}
\begin{split}
  \la R_{1,i}(t) R_{1,j}(t) \ra_{\bfet,\theta}=\int_0^t \diff t' \int_0^t \diff
  t'' e^{-\k \barG
    (t-t')}e^{-\k \barG  (t-t'')}  
\left\la \sum_k \matR_{ik}(t')R_{0,k}(t')
  \sum_l \matR_{jl}(t'')R_{0,l}(t'') \right\ra_{\bfet,\theta}.
\end{split}
\end{equation}
Rearranging and averaging first over the translational noise we get,
\begin{equation}
  \label{eq:correlation_matrix_R2}
\begin{split}
  \la R_{1,i}(t) R_{1,j}(t) \ra_{\bfet,\theta}=\int_0^t \diff t' \int_0^t \diff
  t'' e^{-\k \barG
    (t-t')}e^{-\k \barG  (t-t'')}  
\left\la \sum_{k,l} \matR_{ik}(t') \matR_{jl}(t'') \left\la
    R_{0,k}(t') R_{0,l}(t'')\right\ra_{\bfet}
   \right\ra_{\theta}\\
\end{split}
\end{equation}

\begin{equation}
\la R_{1,i}(t) R_{1,j}(t) \ra_{\bfet,\theta}=2 \kb T \int_0^t \diff t' \int_0^t \diff
  t'' e^{-\k \barG
    (t-t')}e^{-\k \barG  (t-t'')}  
\left\la \sum_{k,l} \matR_{ik}(t') \matR_{jl}(t'') \int_0^{t'} \diff
  t'_1 \int_0^{t''} \diff t'_2 e^{-\k \barG (t'-t'_1)} e^{-\k \barG (t''-t'_2)} \left[ \barG \delta_{kl}
    +\frac{\D \G}{2}\; \matR_{kl}(t'_1) \right] \delta(t'_1 -t'_2)
    \right\ra_{\theta}
\end{equation}

Integrating over the delta function and ignoring the term proportional
to $\D \G$ we get 

\begin{equation}
\begin{split}
\la R_{1,i}(t) R_{1,j}(t) \ra_{\bfet,\theta}&=2 \kb T\barG  e^{-2
\k \barG t}\int_0^t \diff t' \int_0^t \diff
  t'' \int_0^{\min(t',t'')}  \diff t'_1e^{2 \k \barG
      t'_1}
\left\la \sum_{k,l} \matR_{ik}(t') \matR_{jl}(t'') 
  \delta_{kl}
    \right\ra_{\theta} \\
\la R_{1,i}(t) R_{1,j}(t) \ra_{\bfet,\theta}=&2 \kb T\barG  e^{-2
\k \barG t}\int_0^t \diff t' \int_0^t \diff
  t'' \int_0^{\min(t',t'')}  \diff t'_1e^{2 \k \barG
      t'_1}
\left\la \sum_k \matR_{ik}(t') \matR_{jk}(t'')
    \right\ra_{\theta} 
\end{split}
\end{equation}
In order to proceed further, we look at $\la x^2_1(t)
\ra_{\bfet,\theta}$ and $\la y^2_1(t) \ra_{\bfet,\h}$ by setting $i=j$
and subsequently using \cref{eq:eq_5}
\begin{equation}
\begin{split}
\la x^2_1(t) \ra_{\bfet,\theta}=2 \kb T\barG  e^{-2
\k \barG t}\int_0^t \diff t' \int_0^t \diff
  t'' \int_0^{\min(t',t'')}  \diff t'_1e^{2 \k \barG
      t'_1}
\left\la \cos 2[\theta(t')-\theta(t'')] \right\ra_{\theta} 
\end{split}
\end{equation}
Substituting for $\la \cos 2[\h(t')-\h(t'')]\ra_\h$ from
\cref{eq:identity_2} we get
\begin{equation}
\label{eq:x1_sq}
\begin{split}
\la x^2_1(t) \ra_{\bfet,\theta}=2 \kb T\barG  e^{-2
\k \barG t}\int_0^t \diff t' \int_0^t \diff
  t'' \int_0^{\min(t',t'')}  \diff t'_1e^{2 \k \barG
      t'_1}
e^{-4 D_\h \left(t'+t''-2 \min(t',t'')\right)}
\end{split}
\end{equation}


\begin{equation}
\begin{split}
\la x^2_1(t) \ra_{\bfet,\theta}=2 \kb T\barG  e^{-2
\k \barG t}\int_0^t \diff t' \int_0^t \diff
  t'' \frac{e^{2 \k \barG
      \min(t',t'')}-1}{2 \k \barG} e^{-4 D_\h \left(t'+t''-2 \min(t',t'')\right)}
\end{split}
\end{equation}

\begin{equation}
\begin{split}
\la x^2_1(t) \ra_{\bfet,\theta}=\left( \frac{ \kb T}{\k} \right) e^{-2
\k \barG t} \left[\int_0^t \diff t' \int_0^{t'} \diff
  t'' \left(e^{2 \k \barG t''}-1\right) e^{-4 D_\h \left(t'-t''\right)} + \int_0^t \diff t' \int_{t'}^t \diff
  t'' \left(e^{2 \k \barG t'}-1 \right) e^{-4 D_\h \left(t''-t'\right)} \right]
\end{split}
\end{equation}

\begin{equation}
\begin{split}
  \la x^2_1(t) \ra_{\bfet,\theta}=\left( \frac{ \kb T}{\k} \right)
  e^{-2 \k \barG t} \left[\int_0^t \diff t' e^{-4 D_\h t'} \int_0^{t'}
    \diff t'' \left(e^{(2 \k \barG +4 D_\h) t''}- e^{4 D_\h
        t''}\right) + \int_0^t \diff t'\left(e^{2 \k \barG t'}-1
    \right) e^{4 D_\h t'}\int_{t'}^t \diff t'' e^{-4 D_\h t''}
  \right]
\end{split}
\end{equation}

\begin{equation}
\begin{split}
  \la x^2_1(t) \ra_{\bfet,\theta}=\left( \frac{ \kb T}{\k} \right)
  e^{-2 \k \barG t} \left[\int_0^t \diff t' e^{-4 D_\h t'} \left[ \frac{e^{(2 \k \barG +4 D_\h) t'}-1}{2 \k \barG +4 D_\h}- \frac{e^{4 D_\h
        t'}-1}{4 D_\h} \right]+ \int_0^t \diff t'\left(e^{2 \k \barG t'}-1
    \right) e^{4 D_\h t'} \left[  \frac{e^{-4 D_\h t'}-e^{-4 D_\h
          t}}{4D_\h} \right]
  \right]
\end{split}
\end{equation}

\begin{equation}
\begin{split}
  \la x^2_1(t) \ra_{\bfet,\theta}&=\left( \frac{ \kb T}{\k} \right)
  e^{-2 \k \barG t} \left[\int_0^t \diff t' \left( \frac{e^{2 \k \barG  t'}}{2 \k \barG +4 D_\h}-\frac{e^{-4 D_\h t'}}{2 \k \barG +4 D_\h}- \frac{1-e^{-4 D_\h
        t'}}{4 D_\h} \right) \right.\\
&\quad \qquad \qquad \qquad \qquad \qquad \qquad \qquad\left.+ \int_0^t \diff t'\left(\frac{e^{2 \k
        \barG t'}-1}{4 D_\h}\right) -e^{-4D_\h t} \int_0^t \diff t'
 \left(  \frac{e^{(2\k \barG +4 D_\h)t'}-e^{4 D_\h
          t'}}{4D_\h} \right)
  \right]
\end{split}
\end{equation}


\begin{equation}
\begin{split}
  \la x^2_1(t) \ra_{\bfet,\theta}&=\left( \frac{ \kb T}{\k} \right)
  e^{-2 \k \barG t} \left[ \left( \frac{e^{2 \k \barG
          t}-1}{2\k \barG (2 \k \barG +4 D_\h)}-\frac{1-e^{-4 D_\h
          t}}{4 D_\h (2 \k \barG +4 D_\h)}-\frac{t}{4D_\h}+ \frac{1-e^{-4 D_\h
        t}}{16 D^2_\h} \right) \right.\\
&\quad \quad \quad \left.+ \left(\frac{e^{2 \k
        \barG t}-1}{4 D_\h 2 \k \barG}\right) -\frac{t}{4D_\h}-e^{-4D_\h t} 
 \left(  \frac{e^{(2\k \barG +4 D_\h)t}-1}{4D_\h (2 \k \barG +4 D_\h)}- \frac{e^{4 D_\h
          t}-1}{16D^2_\h}\right)
  \right]
\end{split}
\end{equation}

\begin{equation}
\begin{split}
  \la x^2_1(t) \ra_{\bfet,\theta}&=\left( \frac{ \kb T}{\k} \right)
  e^{-2 \k \barG t} \left[ \left( \frac{e^{2 \k \barG
          t}-1}{2\k \barG (2 \k \barG +4 D_\h)}-\frac{1-e^{-4 D_\h
          t}}{4 D_\h (2 \k \barG +4 D_\h)}-\frac{t}{4D_\h}+ \frac{1-e^{-4 D_\h
        t}}{16 D^2_\h} \right) \right.\\
&\quad \quad \quad \left.+ \left(\frac{e^{2 \k
        \barG t}-1}{4 D_\h 2 \k \barG} -\frac{t}{4D_\h}-
  \frac{e^{2\k \barG t}-1}{4D_\h (2 \k \barG +4
     D_\h)}-\frac{1-e^{-4 D_\h t}}{4D_\h (2 \k \barG +4 D_\h)}- \frac{1-e^{-4 D_\h
          t}}{16D^2_\h}\right)
  \right]
\end{split}
\end{equation}

\begin{equation}
\begin{split}
  \la x^2_1(t) \ra_{\bfet,\theta}&=\left( \frac{ \kb T}{\k} \right)
  e^{-2 \k \barG t} \left[ \frac{e^{2 \k \barG
          t}-1}{\k \barG (2 \k \barG +4 D_\h)}-\frac{t}{4D_\h}+\k \barG\frac{1-e^{-4 D_\h
          t}}{4 D^2_\h (2 \k \barG +4 D_\h)}
  \right]
\end{split}
\end{equation}

\begin{equation}
\begin{split}
  \la x^2_1(t) \ra_{\bfet,\theta}&=\left( \frac{ \kb T}{\k} \right)
  \left[ \frac{1-e^{-2 \k \barG t}}{\k \barG (2 \k \barG +4
      D_\h)}-\frac{t e^{-2 \k \barG t}}{4D_\h}+\k \barG\frac{e^{-2 \k
        \barG t}-e^{-(2 \k \barG + 4 D_\h) t}}{4 D^2_\h (2 \k \barG +4 D_\h)} \right]
\end{split}
\end{equation}

\section{Calculation of $\la R_{0,i}(t) R_{2,j}(t) \ra$}
\label{appendix:appendix_4}

\begin{equation}
  \label{eq:eq_7}
\begin{split}
  \la R_{0,i}(t_1) R_{1,j}(t_2) \ra_{\bfet}&=\left\la R_{0,i}(t_1)
    \int_0^{t_2} \diff  t'_2 e^{-\k \barG (t_2 -t'_2)} \sum_k
    \matR_{jk}(t'_2) R_{0,k}(t'_2) \right\ra_{\bfet}\\
  \la R_{0,i}(t_1) R_{1,j}(t_2) \ra_{\bfet}&=
    \int_0^{t_2} \diff  t'_2 e^{-\k \barG (t_2 -t'_2)} \sum_k
    \matR_{jk}(t'_2) \left\la R_{0,i}(t_1)R_{0,k}(t'_2)
    \right\ra_{\bfet}\\
  \la R_{0,i}(t_1) R_{1,j}(t_2) \ra_{\bfet}&=
    \int_0^{t_2} \diff  t'_2 e^{-\k \barG (t_2 -t'_2)} \sum_k
    \matR_{jk}(t'_2) \left[\frac{\kb
    T}{\k} \delta_{ik} \left[e^{-\k \barG (t_1-t'_2)} -e^{-\k \barG
      (t_1+t'_2)} \right] +\kb T\;\D \G e^{-\k \barG
    (t_1+t'_2)}
  \int_0^{\min(t_1,t'_2)} \diff t''  \;e^{2\k \barG t''}
  \matR_{ik}(t'')\right]\\
 \la R_{0,i}(t_1) R_{1,j}(t_2) \ra_{\bfet}&= \left(\frac{\kb
     T}{\k}\right)\;e^{-\k \barG (t_1+t_2)} \int_0^{t_2} \diff  t'_2
 \;e^{\k \barG t'_2} \;\matR_{ji}(t'_2)\left(e^{\k \barG t'_2} -e^{-\k
     \barG t'_2} \right) +\kb T\;\D \G e^{-\k \barG
    (t_1+t_2)} \int_0^{\min(t_1,t'_2)} \diff t''  \;e^{2\k \barG t''} \sum_k
    \matR_{jk}(t'_2) \matR_{ik}(t'')\\
\end{split}
\end{equation}

\begin{equation}
  \label{eq:eq_8}
\begin{split}
   \la R_{0,i}(t_1) R_{1,j}(t_2) \ra_{\bfet,\theta}&= \left(\frac{\kb
     T}{\k}\right)\;e^{-\k \barG (t_1+t_2)} \int_0^{t_2} \diff  t'_2
 \;e^{\k \barG t'_2} \;\left\la \matR_{ji}(t'_2) \right\ra_{\theta}\left(e^{\k \barG t'_2} -e^{-\k
     \barG t'_2} \right) \\
& \qquad \qquad \qquad \qquad \qquad \qquad \qquad \qquad \qquad +\kb T\;\D \G e^{-\k \barG
    (t_1+t_2)} \int_0^{\min(t_1,t'_2)} \diff t''  \;e^{2\k \barG t''} \sum_k
    \left\la\matR_{jk}(t'_2) \matR_{ik}(t'')\right\ra_{\theta}\\
   \la R_{0,i}(t_1) R_{1,j}(t_2) \ra_{\bfet,\theta}&= \left(\frac{\kb
     T}{\k}\right) \matR_{ji}(\h_0)\;e^{-\k \barG (t_1+t_2)} \int_0^{t_2} \diff  t'_2
 \; e^{-4 D_\h t'_2}\left(e^{2\k \barG t'_2} -1 \right) \\
& \qquad \qquad \qquad \qquad \qquad \qquad \qquad \qquad \qquad +\kb T\;\D \G e^{-\k \barG
    (t_1+t_2)} \int_0^{t_2} \diff  t'_2\int_0^{\min(t_1,t'_2)} \diff t''  \;e^{2\k \barG t''} \sum_k
    \left\la\matR_{jk}(t'_2) \matR_{ik}(t'')\right\ra_{\theta}\\
\end{split}
\end{equation}

\begin{equation}
  \label{eq:eq_9}
\begin{split}
 \la x_0(t_1) x_1(t_2) \ra_{\bfet,\theta}&= \left(\frac{\kb
     T}{\k}\right) \cos 2\h_0 \;e^{-\k \barG (t_1+t_2)} \int_0^{t_2} \diff  t'_2
 \; \left(e^{(2\k \barG-4 D_\h) t'_2} -e^{-4 D_\h t'_2} \right) \\
&\qquad \qquad \qquad \qquad \qquad \qquad \qquad \qquad \qquad+\kb T\;\D \G e^{-\k \barG
    (t_1+t_2)}\int_0^{t_2} \diff  t'_2 \int_0^{t'_2} \diff t''  \;e^{2\k \barG t''} \sum_k
    \left\la\matR_{ik}(t'_2) \matR_{ik}(t'')\right\ra_{\theta}\\
 \la x_0(t_1) x_1(t_2) \ra_{\bfet,\theta}&= \left(\frac{\kb
     T}{\k}\right) \cos 2\h_0 \;e^{-\k \barG (t_1+t_2)} \int_0^{t_2} \diff  t'_2
 \; \left(e^{(2\k \barG-4 D_\h) t'_2} -e^{-4 D_\h t'_2} \right) \\
& \qquad \qquad \qquad \qquad \qquad \qquad \qquad \qquad \qquad+\kb T\;\D \G e^{-\k \barG
    (t_1+t_2)}\int_0^{t_2} \diff  t'_2 \int_0^{t'_2} \diff t''  \;e^{2\k \barG t''} \sum_k
    \left\la \cos 2(\h(t'_2)-\h(t'')) \right\ra_{\theta}\\
 \la x_0(t_1) x_1(t_2) \ra_{\bfet,\theta}&= \left(\frac{\kb
     T}{\k}\right) \cos 2\h_0 \;e^{-\k \barG (t_1+t_2)} \int_0^{t_2} \diff  t'_2
 \; \left(e^{(2\k \barG-4 D_\h) t'_2} -e^{-4 D_\h t'_2} \right) \\
& \qquad \qquad \qquad \qquad \qquad \qquad \qquad \qquad \qquad+\kb T\;\D \G e^{-\k \barG
    (t_1+t_2)}\int_0^{t_2} \diff  t'_2 \int_0^{t'_2} \diff t''
  \;e^{2\k \barG t''} e^{-4 D_\h (t'_2+t''-2 \min(t'_2,t''))}\\
 \la x_0(t_1) x_1(t_2) \ra_{\bfet,\theta}&= \left(\frac{\kb
     T}{\k}\right) \cos 2\h_0 \;e^{-\k \barG (t_1+t_2)}  \; \left(\frac{e^{(2\k \barG-4 D_\h) t_2}-1}{2 \k \barG - 4 D_\h}
   -\frac{1-e^{-4 D_\h t_2}}{4 D_\h} \right)+\kb T\;\D \G e^{-\k \barG
    (t_1+t_2)}\int_0^{t_2} \diff  t'_2 \int_0^{t'_2} \diff t''
  \;e^{2\k \barG t''} e^{-4 D_\h (t'_2-t'')}\\
\la x_0(t_1) x_1(t_2) \ra_{\bfet,\theta}&= \left(\frac{\kb
     T}{\k}\right) \cos 2\h_0 \; e^{-\k \barG t_1} \;\left(\frac{e^{-4
       D_\h t_2}-e^{-\k \barG t_2}}{2 \k \barG - 4 D_\h}
   -\frac{e^{-\k \barG t_2}-e^{-(4 D_\h+\k \barG) t_2}}{4 D_\h} \right)+\kb T\;\D \G e^{-\k \barG
    (t_1+t_2)}\int_0^{t_2} \diff  t'_2  \frac{e^{2\k \barG t'_2}
    -e^{-4 D_\h t'_2}}{2\k \barG +4 D_\h}\\
\la x_0(t_1) x_1(t_2) \ra_{\bfet,\theta}&= \left(\frac{\kb
     T}{\k}\right) \cos 2\h_0 \; e^{-\k \barG t_1} \;\left(\frac{e^{(\k
       \barG -4D_\h )t_2}-e^{-\k \barG t_2}}{2 \k \barG - 4 D_\h}
   -\frac{e^{-\k \barG t_2}-e^{-(4 D_\h+\k \barG) t_2}}{4 D_\h}
 \right)\\
&\qquad \qquad \qquad \qquad \qquad \qquad \qquad \qquad
\qquad+\kb T\;\D \G
 e^{-\k \barG (t_1+t_2)}\left[\frac{e^{2\k \barG t_2}-1}{2 \k \barG (2\k \barG +4 D_\h)}
    -\frac{1-e^{-4 D_\h t_2}}{4D_\h(2\k \barG +4 D_\h)}\right]\\
\end{split}
\end{equation}

\begin{equation}
\label{eq:eq_10}
\begin{split}
\la x_0(t_1) x_1(t_2) \ra_{\bfet,\theta}&= \left(\frac{\kb
     T}{\k}\right) \cos 2\h_0 \; e^{-\k \barG t_1} \;\left(\frac{e^{(\k
       \barG -4D_\h )t_2}-e^{-\k \barG t_2}}{2 \k \barG - 4 D_\h}
   -\frac{e^{-\k \barG t_2}-e^{-(4 D_\h+\k \barG) t_2}}{4 D_\h}
 \right)\\
&\qquad \qquad \qquad \qquad \qquad \qquad \qquad \qquad
\qquad+\left(\frac{\kb T}{\k}\right)\;\left(\frac{\D \G}{2\barG}\right)
 e^{-\k \barG t_1}\left[\frac{e^{\k \barG t_2}-e^{-\k \barG t_2}}{(2\k \barG +4 D_\h)}
    -\frac{2\k\barG}{4D_\h}\frac{e^{-\k \barG t_2}-e^{-(2 \k \barG +4
        D_\h) t_2}}
{(\k \barG +4 D_\h)}\right]\\
\end{split}
\end{equation}

\begin{equation}
  \label{eq:eq_11}
\begin{split}
  \left\la R_{0,i}(t) R_{2,j}(t) \right\ra_{\bfet,\theta}=&\left \la
    R_{0,i}(t) \int_0^t \diff t' e^{-\k \barG (t-t')} \sum_k
    \matR_{jk} (t') R_{1,k}(t') \right \ra_{\bfet,\theta}=\int_0^t
  \diff t' e^{-\k \barG (t-t')} \left \la\sum_k
    \matR_{jk} (t') \left \la
    R_{0,i}(t) R_{1,k}(t') \right \ra_{\bfet} \right \ra_\h\\
\end{split}
\end{equation}

\begin{equation}
  \label{eq:eq_12}
  \begin{split}
    \la R_{0,i}(t) R_{1,k}(t') \ra_{\bfet}&= \left(\frac{\kb
     T}{\k}\right)\;e^{-\k \barG (t+t')} \int_0^{t'} \diff  t'_2
 \;e^{\k \barG t'_2} \;\matR_{ki}(t'_2)\left(e^{\k \barG t'_2} -e^{-\k
     \barG t'_2} \right) +\kb T\;\D \G e^{-\k \barG
    (t+t')} \int_0^{t'} \diff t''  \;e^{2\k \barG t''} \sum_l
    \matR_{kl}(t'_2) \matR_{il}(t'')\\ 
  \end{split}
\end{equation}
Neglecting the second term in \cref{eq:eq_12}, we have
\begin{equation}
  \label{eq:eq_13}
\begin{split}
  \left\la R_{0,i}(t) R_{2,j}(t) \right\ra_{\bfet,\theta}&=\int_0^t
  \diff t' e^{-\k \barG (t-t')} \left \la\sum_k
    \matR_{jk} (t') \left(\frac{\kb
     T}{\k}\right)\;e^{-\k \barG (t+t')} \int_0^{t'} \diff  t'_2
 \;e^{\k \barG t'_2} \;\matR_{ki}(t'_2)\left(e^{\k \barG t'_2} -e^{-\k
     \barG t'_2} \right) \right \ra_\h\\
  \left\la R_{0,i}(t) R_{2,j}(t) \right\ra_{\bfet,\theta}&=\left(\frac{\kb
     T}{\k}\right)\int_0^t
  \diff t' \int_0^{t'} \diff  t'_2 e^{-\k \barG (t-t')} \;e^{-\k \barG
    (t+t')}  \;e^{\k \barG t'_2} \left(e^{\k \barG t'_2} -e^{-\k
     \barG t'_2} \right)  \left \la\sum_k
    \matR_{jk} (t')  \;\matR_{ki}(t'_2) \right\ra_\h\\
 \left\la R_{0,i}(t) R_{2,j}(t) \right\ra_{\bfet,\theta}&=\left(\frac{\kb
     T}{\k}\right)e^{-2 \k \barG t}\int_0^t
  \diff t' \int_0^{t'} \diff  t'_2  \left(e^{2\k \barG t'_2} -1 \right)  \left \la\sum_k
    \matR_{jk} (t')  \;\matR_{ki}(t'_2) \right\ra_\h\\
\end{split}
\end{equation}

For the mean-square displacement along $x$ and $y$ direction, setting
$j=i$ and using \cref{eq:eq_5} we get
\begin{equation}
  \label{eq:eq_14}
  \begin{split}
 \left\la x_0(t) x_2(t) \right\ra_{\bfet,\theta}&=\left\la y_0(t) y_2(t) \right\ra_{\bfet,\theta}=\left(\frac{\kb
     T}{\k}\right)e^{-2 \k \barG t}\int_0^t
  \diff t' \int_0^{t'} \diff  t'_2  \left(e^{2\k \barG t'_2} -1
  \right)  \left \la \cos 2 \left(\h(t') -\h(t'_2)\right)
  \right\ra_\h\\  
&=\left(\frac{\kb
     T}{\k}\right)e^{-2 \k \barG t}\int_0^t
  \diff t' \int_0^{t'} \diff  t'_2  \left(e^{2\k \barG t'_2} -1
  \right) e^{-4 D_\h (t'-t'_2)}\\  
&=\left(\frac{\kb
     T}{\k}\right)e^{-2 \k \barG t}\int_0^t
  \diff t'  e^{-4 D_\h t'} \int_0^{t'} \diff  t'_2  \left(e^{(2\k
      \barG+4 D_\h) t'_2} -e^{4 D_\h t'_2}
  \right) \\  
&=\left(\frac{\kb
     T}{\k}\right)e^{-2 \k \barG t}\int_0^t
  \diff t'  e^{-4 D_\h t'}  \left(\frac{e^{(2\k
      \barG+4 D_\h) t'}-1}{2\k\barG+4 D_\h} -\frac{e^{4 D_\h t'}-1}{4 D_\h}
  \right) \\  
&=\left(\frac{\kb
     T}{\k}\right)e^{-2 \k \barG t}\int_0^t
  \diff t'    \left(\frac{e^{2\k
      \barG t'}-e^{-4 D_\h t'}}{2\k\barG+4 D_\h} -\frac{1-e^{-4 D_\h t'}}{4 D_\h}
  \right) \\  
&=\left(\frac{\kb
     T}{\k}\right)e^{-2 \k \barG t}\left(\frac{e^{2\k
      \barG t}-1}{2 \k \barG (2\k\barG+4 D_\h)}-\frac{1-e^{-4 D_\h
      t}}{4 D_\h(2\k\bar+4 D_\h)} -\frac{t}{4 D_\h}-\frac{1-e^{-4 D_\h t'}}{16 D^2_\h}
  \right) \\  
&=\left(\frac{\kb
     T}{\k}\right)\left[\frac{1-e^{-2\k
      \barG t}}{2 \k \barG (2\k\barG+4 D_\h)}-\frac{te^{-2\k \barG t}
  }{4 D_\h}+\frac{2\k \barG}{4 D_\h}\left(1-e^{-4 D_\h
      t}\right)  \right] \\  
  \end{split}
\end{equation}

\end{appendices}

\end{document}